\documentclass{aastex}

\newcommand{\Msun}{\mbox{M$_{\odot}$}}
\newcommand{\Lsun}{\mbox{L$_{\odot}$}}
\newcommand{\arcminspace}{\mbox{$\arcmin$ }}
\newcommand{\arcsecspace}{\mbox{$\arcsec$ }}

\usepackage{color}

\begin{document}

\title{Young stellar clusters containing massive Young Stellar Objects in the VVV Survey.}

\author{J.\ Borissova\altaffilmark{1,2}, 
S.\ Ram\'{\i}rez Alegr\'{\i}a\altaffilmark{1,2}, 
J.\ Alonso\altaffilmark{2,3}, 
P.~W. Lucas\altaffilmark{4}, 
R.\ Kurtev\altaffilmark{1,2}, 
N.\ Medina\altaffilmark{1,2}, 
C.\ Navarro\altaffilmark{1,2}, 
M.\ Kuhn\altaffilmark{1,2}, 
M.\ Gromadzki\altaffilmark{1,2}, 
G.\ Retamales\altaffilmark{1,2},  
M.~A.\ Fernandez\altaffilmark{1,2}, 
C.\ Agurto-Gangas\altaffilmark{1,2}, 
A.-N.\ Chen\'e\altaffilmark{5},  
D.\ Minniti\altaffilmark{2,6,7}, 
C.\ Contreras Pe\~na\altaffilmark{4}, 
M. Catelan\altaffilmark{2,3}, 
I.\ Decany\altaffilmark{2,3}, 
M.~A.\ Thompson\altaffilmark{4}, 
E.\ F.\ E.\ Morales\altaffilmark{8}, 
P.\ Amigo\altaffilmark{2,2}}
\affil{$^1$Instituto de F\'isica y Astronom\'ia, Universidad de Valpara\'iso, Av. Gran Breta\~na 1111, Playa Ancha, Casilla 5030, Chile}
\authoremail{jura.borissova@uv.cl}
\affil{$^2$Millennium Institute of Astrophysics (MAS), Santiago, Chile} 
\affil{$^3$Instituto de Astrof\'isica, Facultad de F\'isica, Pontificia Universidad Cat\'olica de Chile, Casilla 306, Santiago 22, Chile}
\affil{$^4$Centre for Astrophysics Research, Science and Technology Research Institute, University of Hertfordshire, Hatfield AL10 9AB, UK}          
\affil{$^5$Gemini Observatory, Northern Operations Center, 670 N. A'ohoku Place, Hilo, HI 96720, USA}                  
\affil{$^6$Departamento de Ciencias F\'isicas, Universidad Andres Bello, Republica 220, Santiago, Chile}   
\affil{$^7$Vatican Observatory, V00120 Vatican City State, Italy}                                          
\affil{$^8$Max-Planck-Institute for Astronomy, Germany} 

\received{2015 May 23}
\accepted{2016 June 17}

\begin{abstract}

The purpose of this research is to study the connection of global properties of eight young stellar clusters projected in the Vista Variables in the Via Lactea (VVV) ESO Large Public Survey disk area and their young stellar object population. The analysis in based on the combination of spectroscopic parallax-based reddening and distance determinations with main sequence and pre-main sequence ishochrone fitting to determine the basic parameters (reddening, age, distance) of the sample clusters. The lower mass limit estimations show that all clusters are low or intermediate mass (between 110 and 1800 \Msun), the slope $\Gamma$ of the obtained present-day mass functions of the clusters is close to the Kroupa initial mass function. On the other hand, the young stellar objects in the surrounding cluster's fields are classified by low resolution spectra, spectral energy distribution fit with theoretical predictions and variability, taking advantage of multi-epoch VVV observations.  All spectroscopically confirmed young stellar objects (except one) are found to be  massive (more than 8 \Msun). Using VVV and GLIMPSE color-color cuts we have selected  a large number of new young stellar object candidates, which are checked for variability and  57 \% are found to show at least low-amplitude variations. In few cases it was possible to distinguish between YSO and AGB classification on the basis of the light curves.
 
\end{abstract}


\section{Introduction}

The Vista Variables in the V\'{\i}a L\'actea (VVV) is one of the six ESO Public Surveys using the 4 m VISTA telescope (Arnaboldi et al. 2007), which scans the Galactic Bulge and southern Disk in five near-infrared filters (Minniti et al. 2010; Saito et al. 2010; Saito et al. 2012). The VVV data are publicly available through the VISTA Science Archive (VSA; Cross et al. 2012). Technical information about the survey can be found in Saito et al. (2012) and Soto et al. (2013). A primary goal of the VVV is to describe the numerous star clusters in its coverage area in detail, which is made possible by the infrared nature of the VVV survey, its small pixel size, and its depth, which reduce the influence of dust absorption and nebulosity in the crowded regions of the Galactic Plane. In Borissova et al. (2011), we presented a catalog of 96 new cluster candidates in the disk area covered by the VVV survey. In Chen\'{e} et al. (2012) we described the methodology employed to establish cluster parameters by analyzing four known young clusters: Danks\,1, Danks\,2, RCW\,79, and [DBS2003]\,132. In Chen\'{e} et al. (2013, 2015) we presented the first study of seven clusters from the Borissova et al. (2011) catalog, which contains at least one newly discovered Wolf-Rayet (WR) star, member of these clusters. Later on, we use the radiative transfer code CMFGEN to analyze the K-band spectra of these stars and to derive the stellar parameters and surface abundances for a subset of them (Herv\'{e} et~al. 2016). 
In Ram{\'{\i}}rez Alegr{\'{\i}}a et  al. (2014)  we presented the physical characterization of VVV\,CL086, a new massive cluster, found at the far end of the Milky Way bar at a distance of 11$\pm6$ kpc, and finally in Borissova et al. (2014) we reported the results of our search for new star cluster candidates projected on the inner disk and bulge area covered by the VVV survey. 

In this paper we continue with analysis of star clusters, using the VVV database. We present eight Galactic young clusters, which contain Young Stellar Objects (YSOs) and/or stars with emission lines in their spectra. The main goal of this investigation is to combine photometric data for the clusters with spectroscopy of individual YSOs to better determine the properties of both the clusters and the individual YSOs.
 
\section{The sample}

\begin{figure}
\epsscale{1.0}
\plotone{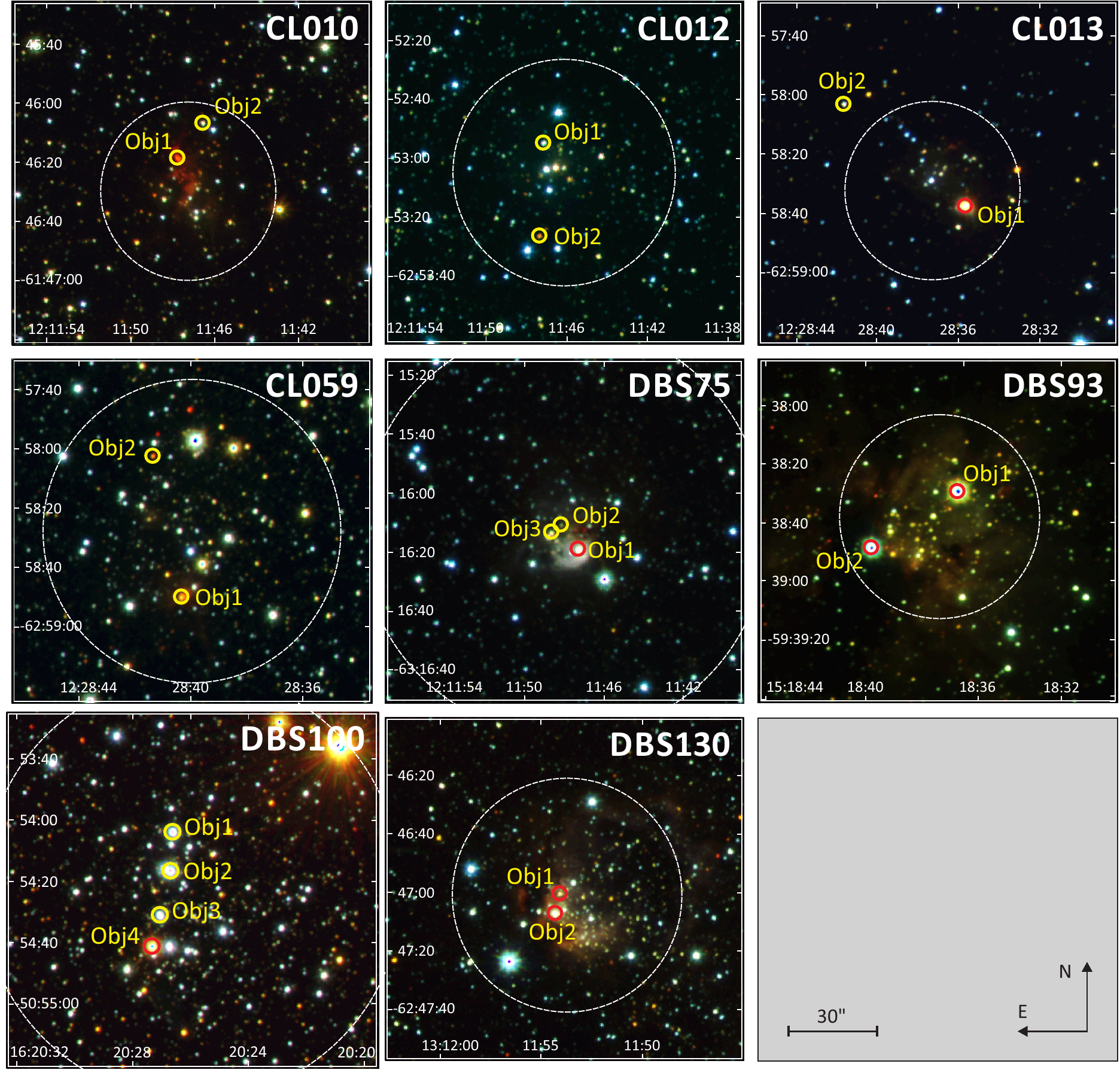}
\vspace{0.2cm}
\caption{VVV $JHK_{\rm S}$ composite color images of VVV\,CL010, VVV\,CL012, VVV\,CL013, VVV\,CL059, [DBS2003]\,75, [DBS2003]\,93, [DBS2003]\,100, and [DBS2003]\,130. North is up, east to the left. The stars with emission lines in their spectra are marked.} 
\label{rgb_all}
\end{figure}

Four of the clusters in our sample are selected from the Borissova et al. (2011) catalog and the rest of the clusters are selected from Dutra et al. (2003) catalog. All of them are in the VVV disk area (Minniti et al. 2010, Saito et al. 2012), and the $J$, $H$, $K_S$ near-infrared images are used to construct the three-color images (Fig.~\ref{rgb_all}) and for photometric analysis. The coordinates of the clusters are given in Table~\ref{main_par}. For the VVV clusters, this is the first time when the photometric and spectroscopic analysis is presented. The [DBS2003]\,130 is investigated by Baume et al. (2009) and $E(B-V)$ of 2.3 mag and age of 1-2 Myr were determined using the combination between optical and near-infrared photometry; the cluster [DBS2003]\,93 was mentioned as a small embedded cluster, inside the GAL322.16+00.62 H{\sc ii} region by Moises et al. (2011), but no deep color-magnitude diagram was reported. The clusters [DBS2003]\,75, [DBS2003]\,93,  [DBS2003]\,100 are analyzed by Kharchenko et al. (2013) team using the 2MASS photometry (which has limiting magnitude around 14 in $K_S$ band). 

\begin{deluxetable}{lllllllllll}
\tabletypesize{\tiny}
\setlength{\tabcolsep}{0.08in} 
\tablecaption{Main parameters of the clusters in the sample.}
\tablehead{
\colhead{Name}            &
\colhead{RA}  &
\colhead{DEC}    &
\colhead{l}    &
\colhead{b} &
\colhead{E(J-K)} &
\colhead{$(M-m)_0$}&
\colhead{Age$/$Myr}&
\colhead{Mass/M$_{\odot}$}&
\colhead{$\Gamma$}&
\colhead{Radius}\arcminspace}
\startdata
VVV\,CL010 \footnote{The distances derived by different methods yield very discrepant results, thus the value reported in the Table is uncertain.} &12:11:47&$-$61:46:24&298.261&$+$0.738&2.14$\pm$0.2&14.55$\pm$1.3&2.7$\pm$1.5&(1.8$\pm$0.5)$\cdot$10$^3$&-0.88$\pm$0.15 & 30$\pm$9 \\
VVV\,CL012&12:20:14&$-$62:53:06&299.385&$-$0.228&2.0$\pm$0.3&14.1$\pm$0.7&10.0$\pm$2.5&(1.1$\pm$0.4)$\cdot$10$^2$&-1.31$\pm$0.25& 37$\pm$3 \\
VVV\,CL013&12:28:37&$-$62:58:24&300.343&$-$0.216&2.1$\pm$0.3&13.2$\pm$1.1&3.0$\pm$2.0&(1.3$\pm$0.3)$\cdot$10$^2$&-1.33$\pm$0.08& 30$\pm$3 \\
VVV\,CL059&16:05:52&$-$50:47:48&331.243&$+$1.067&3.0$\pm$0.2&15.3$\pm$1.4&20.0$\pm$3.2&(8.3$\pm$2.1)$\cdot$10$^2$&-1.09$\pm$0.24& 55$\pm$8 \\
$[$DBS2003$]$\,75&12:09:02&$-$63:15:54&298.184&$-$0.785&1.5$\pm$0.2&13.6$\pm$1.3&2.0$\pm$1.0&(8.6$\pm$1.8)$\cdot$10$^2$&-1.00$\pm$0.12& 72$\pm$3 \\
$[$DBS2003$]$\,93&15:18:37&$-$56:38:42&322.160&$+$0.629&2.6$\pm$0.3&11.6$\pm$0.9&20.0$\pm$5.0&(2.5$\pm$0.8)$\cdot$10$^2$&-0.95$\pm$0.09& 36$\pm$4\\       
$[$DBS2003$]$\,100&16:20:26&$-$50:54:24&332.840&$-$0.590&1.1$\pm$0.1&12.78$\pm$0.8&12.0$\pm$2.5&(5.2$\pm$1.1)$\cdot$10$^2$&-1.18$\pm$0.09&80$\pm$12\\
$[$DBS2003$]$\,130&13:11:54&$-$62:47:00&305.269&$-$0.004&2.5$\pm$0.2&13.1$\pm$1.2&3.0$\pm$1.0&(2.2$\pm$0.9)$\cdot$10$^2$&-1.13$\pm$0.34&41$\pm$11\\
\enddata
\label{main_par}
$^1$ The distances derived by different methods yield very discrepant results, thus the value reported in the Table is uncertain.
\end{deluxetable}

\section{The color-magnitude diagrams and fundamental parameters of the clusters.}

The procedure employed for determining the fundamental cluster parameters such as age, reddening, and distance is described in Borissova et al. (2011, 2014) and Chen\'{e} et al. (2012, 2013). Briefly, to construct the color-magnitude diagram we perform a PSF photometry of $10\times10$ arcmin J, H, and $K_S$ fields surrounding the selected candidate. Each image was taken with 80 sec exposure time. We used the VVV-SkZ pipeline, which is an automatic PSF-fitting photometric pipeline for the VVV survey (Mauro et al. 2013)  and Dophot (Alonso Garcia et al. 2015). The saturated stars (usually $K_S \leq 11.5$ mag, depending from the crowding) were replaced by 2\,MASS stars (Point Source Catalog).  Since 2\,MASS has a much lower angular resolution than the VVV, when replacing stars we carefully examined each cluster to avoid contamination effects of crowding, using the Point Source Catalog Quality Flags given in the 2MASS catalog. To separate the field stars from probable cluster members we used the latest version of the field-star decontamination algorithm of Bonatto \& Bica (2010). The algorithm divides the $K_S$ , (H - $K_S$) and (J - $K_S$) ranges into a grid of cells. In each cell, it estimates the expected number density of cluster stars by subtracting the respective field-star number density and, summing over all cells, it obtains a total number of member stars. Grid shifts of $\pm 1/3$ the cell size are applied in each axes. The average of these is the limit for considering a star as a possible cluster member. Only the highest survival frequency after all tests were finally considered as cluster members.

We have collected spectra of 16 stars (Table~\ref{main_par_stars}) using the IR spectrograph and imaging camera SofI in long-slit mode, mounted on the ESO New Technology Telescope (NTT), the Infrared Spectrometer and Array Camera (ISAAC) mounted on the VLT, and the Ohio State InfraRed Imager/Spectrometer (OSIRIS) mounted on the Southern Observatory for Astrophysical Research (SOAR) telescope.\footnote{{Based on observations gathered with VIRCAM, VISTA of the ESO as part of observing programs 179.B-2002; ESO programs 087.D-0341(A); 087.D-0490(A) and 089.D-0462(A)}}
 The instrument set-ups give resolution of R=2200 for SofI; 3000 for ISAAC; and 3000 for OSIRIS. Total exposure times were typically 200--400\,s for the brightest stars and 1200\,s for the faintest. The reduction procedure for the spectra is described in Chen\'{e} et al. (2012, 2013). Spectral classification was performed using atlases of $K$-band spectra that feature spectral types stemming from optical studies (Rayner et al. 2009; Hanson et al. 1996, 2005) in concert with the spectral atlases of Martins \& Coelho (2007), Crowther et al. (2006), Liermann et al. (2009),  Mauerhan et al. (2011), Meyer et al. (1998), and  Wallace \& Hinkle (1997). The equivalent widths (EWs) were measured from the continuum-normalized spectra using the {\sc iraf}\footnote{IRAF is distributed by the National Optical Astronomy Observatory, which is operated by the Association of Universities for Research in Astronomy (AURA) under cooperative agreement with the National Science Foundation.} task {\it splot}. When the S/N was high enough, the luminosity class of the star was determined using the EW of the CO line and the Davies et al. (2007) calibration. However, for spectroscopic targets displaying low S/N it was difficult to distinguish luminosity Class~I objects from their Class~III counterparts. Individual extinction and distance was estimated using the spectral classifications of the objects and the intrinsic colors and luminosities cited by Martins \& Plez (2006) for O type stars and by Strai\v{z}ys \& Lazauskait\.{e} (2009) for the rest of the spectral types (tabulated in Table~\ref{main_par_stars1}). The uncertainties are calculated by quadratically adding the uncertainties of the photometry and the spectral classification (e.g., 2 subtypes). 

\begin{deluxetable}{llllllllllllll}
\tabletypesize{\tiny} \rotate
\setlength{\tabcolsep}{0.02in} 
\tablewidth{0pt}
\tablecaption{Main parameters of the observed stars in the sample.}
\tablehead{
\colhead{Name}            &
\colhead{RA}  &
\colhead{DEC}    &
\colhead{J}    &
\colhead{H} &
\colhead{K}&
\colhead{3.6}&
\colhead{4.5}&
\colhead{5.8}    &
\colhead{8.0} &
\colhead{W1}&
\colhead{W2}&
\colhead{W3}&
\colhead{W4}\\
\colhead{(1)}&
\colhead{(2)}&
\colhead{(3)}&
\colhead{(4)}&
\colhead{(5)}&
\colhead{(6)}&
\colhead{(7)}&
\colhead{(8)}&
\colhead{(9)}&
\colhead{(10)}&
\colhead{(11)}&
\colhead{(12)}&
\colhead{(13)}&
\colhead{(14)}
}
\startdata
  CL010 Obj1  & 182.948679 & -61.771908 & 17.18$\pm$0.20 & 14.87$\pm$0.14 & 11.07$\pm$0.20 &  8.37$\pm$0.30 &   \nodata      &  6.67$\pm$0.20 &   \nodata      &   8.41$\pm$0.02 &  6.16$\pm$0.02 &  3.62$\pm$0.01 & -1.71$\pm$0.01\\
  CL010 Obj2  & 182.943884 & -61.768623 & 14.74$\pm$0.06 & 13.57$\pm$0.04 & 12.72$\pm$0.08 & 11.69$\pm$0.05 & 11.14$\pm$0.08 & 10.84$\pm$0.08 & 10.43$\pm$0.06 &    \nodata      &   \nodata      &   \nodata      &   \nodata     \\
  CL012 Obj1  & 185.062471 & -62.882065 & 14.39$\pm$0.05 & 13.21$\pm$0.05 & 12.42$\pm$0.07 &   \nodata      &   \nodata      &   \nodata      &   \nodata      &    \nodata      &   \nodata      &   \nodata      &   \nodata     \\
  CL012 Obj2  & 185.063708 & -62.890823 & 14.64$\pm$0.30 & 14.01$\pm$0.30 & 13.51$\pm$0.07 & 10.37$\pm$0.04 &  9.26$\pm$0.06 &  8.36$\pm$0.03 &  7.52$\pm$0.04 &   9.71$\pm$0.03 &  8.28$\pm$0.02 &  5.10$\pm$0.02 &  1.81$\pm$0.03\\
  CL013 Obj1  & 187.148962 & -62.976452 & 13.29$\pm$0.06 & 10.73$\pm$0.04 &  8.68$\pm$0.03 &  6.77$\pm$0.12 &   \nodata      &  4.17$\pm$0.03 &   \nodata      &   5.89$\pm$0.05 &  4.34$\pm$0.06 &  1.23$\pm$0.02 & -1.01$\pm$0.01\\
  CL013 Obj2  & 187.153946 & -62.973420 & 15.18$\pm$0.02 & 13.92$\pm$0.02 & 12.96$\pm$0.02 &   \nodata      &   \nodata      &   \nodata      &   \nodata      &    \nodata      &   \nodata      &   \nodata      &   \nodata     \\
  CL013 Obj3  & 187.156037 & -62.974136 & 13.01$\pm$0.05 & 12.30$\pm$0.08 & 11.62$\pm$0.06 &   \nodata      &   \nodata      &   \nodata      &   \nodata      &  11.00$\pm$0.03 & 10.37$\pm$0.03 &  7.69$\pm$0.17 &  6.06$\pm$0.15\\
  CL013 Obj4  & 187.174247 & -62.967098 & 14.15$\pm$0.03 & 13.04$\pm$0.03 & 12.10$\pm$0.02 & 10.82$\pm$0.05 & 10.26$\pm$0.05 &  9.73$\pm$0.05 &  8.71$\pm$0.04 &   6.40$\pm$0.06 &  6.09$\pm$0.04 &  2.78$\pm$0.03 &  0.44$\pm$0.01\\
  CL059 Obj1  & 241.467897 & -50.802849 & 15.34$\pm$0.30 & 13.17$\pm$0.07 & 10.46$\pm$0.04 &  7.63$\pm$0.14 &  6.54$\pm$0.11 &  5.52$\pm$0.05 &  4.89$\pm$0.08 &   7.40$\pm$0.03 &  5.29$\pm$0.03 &  1.33$\pm$0.01 & -1.61$\pm$0.02\\
  CL059 Obj2  & 241.472292 & -50.789556 & 17.66$\pm$0.06 & 15.03$\pm$0.01 & 12.43$\pm$0.26 & 11.07$\pm$0.11 &  9.90$\pm$0.06 &  9.04$\pm$0.08 &  8.32$\pm$0.11 &   9.52$\pm$0.06 &  8.08$\pm$0.03 &  4.68$\pm$0.02 &  0.70$\pm$0.03\\
  DBS75 Obj1  & 182.255295 & -63.266605 & 11.79$\pm$0.08 & 10.53$\pm$0.07 &  9.10$\pm$0.06 &   \nodata      &   \nodata      &   \nodata      &   \nodata      &   4.08$\pm$0.08 &  4.01$\pm$0.03 & -1.33$\pm$0.13 & -4.18$\pm$0.01\\
  DBS75 Obj2  & 182.258865 & -63.264286 & 15.80$\pm$0.01 & 14.51$\pm$0.01 & 13.27$\pm$0.01 &   \nodata      &   \nodata      &   \nodata      &   \nodata      &    \nodata      &   \nodata      &   \nodata      &   \nodata     \\
  DBS75 Obj3  & 182.260925 & -63.265038 & 14.45$\pm$0.01 & 12.75$\pm$0.01 & 11.29$\pm$0.01 &   \nodata      &   \nodata      &   \nodata      &   \nodata      &    \nodata      &   \nodata      &   \nodata      &   \nodata     \\
  DBS93 Obj1  & 229.650547 & -56.641270 &  9.28$\pm$0.02 &  7.63$\pm$0.02 &  6.93$\pm$0.02 &  6.76$\pm$0.04 &  6.76$\pm$0.06 &  6.41$\pm$0.03 &   \nodata      &   6.52$\pm$0.96 &  6.80$\pm$0.30 & -0.54$\pm$0.35 & -5.44$\pm$0.50\\
  DBS93 Obj2  & 229.665245 & -56.646664 &  8.75$\pm$0.02 &  8.20$\pm$0.02 &  8.07$\pm$0.03 &  7.92$\pm$0.06 &  8.03$\pm$0.16 &   \nodata      &   \nodata      &    \nodata      &   \nodata      &   \nodata      &   \nodata     \\
 DBS100 Obj1  & 245.111029 & -50.901295 & 10.05$\pm$0.04 &  9.44$\pm$0.03 &  9.15$\pm$0.03 &  8.88$\pm$0.04 &  8.86$\pm$0.05 &  8.82$\pm$0.04 &  9.01$\pm$0.11 &   8.64$\pm$0.06 &  8.54$\pm$0.06 &  5.39$\pm$0.30 &  0.80$\pm$0.23\\
 DBS100 Obj2  & 245.111599 & -50.904770 &  9.49$\pm$0.04 &  8.92$\pm$0.04 &  8.58$\pm$0.03 &   \nodata      &   \nodata      &   \nodata      &   \nodata      &   8.15$\pm$0.04 &  8.15$\pm$0.06 &  6.13$\pm$0.58 &  1.34$\pm$0.20\\
 DBS100 Obj3  & 245.112948 & -50.908836 & 10.61$\pm$0.03 &  9.95$\pm$0.03 &  9.69$\pm$0.03 &   \nodata      &   \nodata      &   \nodata      &   \nodata      &    \nodata      &   \nodata      &   \nodata      &   \nodata     \\
 DBS100 Obj4  & 245.114143 & -50.911667 & 11.66$\pm$0.04 &  9.12$\pm$0.04 &  7.94$\pm$0.02 &   \nodata      &   \nodata      &   \nodata      &   \nodata      &    \nodata      &   \nodata      &   \nodata      &   \nodata     \\
 DBS130 Obj1  & 197.975786 & -62.783649 & 14.98$\pm$0.30 & 13.02$\pm$0.18 & 12.44$\pm$0.15 &   \nodata      &   \nodata      &   \nodata      &   \nodata      &    \nodata      &   \nodata      &   \nodata      &   \nodata     \\
 DBS130 Obj2  & 197.976772 & -62.785511 & 12.51$\pm$0.05 & 10.87$\pm$0.06 &  9.81$\pm$0.04 &   \nodata      &   \nodata      &   \nodata      &   \nodata      &    \nodata      &   \nodata      &   \nodata      &   \nodata     \\
\enddata
\label{main_par_stars}
\end{deluxetable}

\begin{deluxetable}{llllllllll}
\tabletypesize{\tiny} 
\setlength{\tabcolsep}{0.01in} 
\tablewidth{0pt}
\tablecaption{Main parameters of the observed stars in the sample and Log of observations.}
\tablehead{
\colhead{Name}            &
\colhead{S70(Jy)}&
\colhead{S160(Jy)}&
\colhead{S250(Jy)}&
\colhead{S350(Jy)}&
\colhead{S500(Jy)}&
\colhead{Sp.type}&
\colhead{M$_k$}&
\colhead{(J-K)$_0$}&
\colhead{Log}\\
\colhead{(1)}&
\colhead{(15)}&
\colhead{(16)}&
\colhead{(17)}&
\colhead{(18)}&
\colhead{(19)}&
\colhead{(20)}&
\colhead{(21)}&
\colhead{(22)}&
\colhead{(23)}}           
\startdata
  CL010 Obj1  &    \nodata       &    \nodata       & 247.61$\pm$ 6.83 & 86.34$\pm$1.96 & 39.42$\pm$1.03 & \nodata   & \nodata & \nodata&  2012-05-05T00:33:57.1361, SofI \\
  CL010 Obj2  &    \nodata       &    \nodata       &    \nodata       &   \nodata      &   \nodata      &  O6-8  &-3.86  &-0.21  &  2012-05-05T00:33:57.1361, SofI \\
  CL012 Obj1  &    \nodata       &    \nodata       &    \nodata       &   \nodata      &   \nodata      &  B1-2  & -2.25& -0.13 &  2010-05-03T23:50:21.175, SOAR \\
  CL012 Obj2  &    \nodata       &    \nodata       &    \nodata       &   \nodata      &   \nodata      &  \nodata  &  \nodata &  \nodata &  no spectra, literature YSO cand. \\
  CL013 Obj1  & 136.16$\pm$10.04 &  72.81$\pm$ 6.01 &  20.31$\pm$ 7.39 & 14.17$\pm$7.43 &  0.00$\pm$ \nodata &  O8-B0  &-3.28  & -0.21 &  2010-05-03T00:20:35.271, SOAR \\
  CL013 Obj2  &    \nodata       &    \nodata       &    \nodata       &   \nodata      &   \nodata      &  Be & \nodata & \nodata &  2012-05-06T03:21:08.2385, SofI \\
  CL013 Obj3  &    \nodata       &    \nodata       &    \nodata       &   \nodata      &   \nodata      &  B2-3V  & -1.66  & -0.11 &  2012-05-06T03:21:08.2385, SofI \\
  CL013 Obj4  &    \nodata       &    \nodata       &    \nodata       &   \nodata      &   \nodata      &  \nodata  & \nodata  & \nodata  &  no spectra, literature YSO cand. \\
  CL059 Obj1  & 129.37$\pm$ 2.05 & 125.44$\pm$11.58 & 151.72$\pm$ 6.60 & 62.34$\pm$3.58 & 29.34$\pm$2.76 &  &  \nodata &  \nodata  &  2011-04-11T06:46:40.0196, ISAAC \\
  CL059 Obj2  &    \nodata       &    \nodata       &    \nodata       &   \nodata      &   \nodata  &  \nodata    &  \nodata  &  \nodata  &  no spectra, literature YSO cand. \\
  DBS75 Obj1  & 693.44$\pm$33.09 &   0.00: &    \nodata       &   \nodata  &   \nodata    &  O7-B0   & -3.86 &-0.21 &  2011-04-17T00:20:50.0068, SofI \\
  DBS75 Obj2  &    \nodata       &    \nodata       &    \nodata       &   \nodata      &   \nodata      &  \nodata &  \nodata &  \nodata  &  no spectra, literature YSO cand. \\
  DBS75 Obj3  &    \nodata       &    \nodata       &    \nodata       &   \nodata      &      \nodata    &  \nodata  &   \nodata  &   \nodata &  no spectra, literature YSO cand. \\
  DBS93 Obj1  &    \nodata       &    \nodata       &    \nodata       &   \nodata      &   \nodata      &  \nodata  &  \nodata &  \nodata  &  2011-04-15T03:35:36.6140, SofI \\
  DBS93 Obj2  & 844.40$\pm$28.34 & 975.70: &    \nodata       &   \nodata      &   \nodata      &  \nodata  &  \nodata &  \nodata &  2011-04-15T03:35:36.6140, SofI \\
 DBS100 Obj1  &    \nodata       &    \nodata       &    \nodata       &   \nodata      &   \nodata      &   O4-5  & -4.68 & -0.21	 &  2011-04-17T08:52:41.0839, SofI \\
 DBS100 Obj2  &    \nodata       &    \nodata       &    \nodata       &   \nodata      &   \nodata      &   O7-8  & -3.86 & -0.21	 &  2011-04-17T08:52:41.0839, SofI \\
 DBS100 Obj3  &    \nodata       &    \nodata       &    \nodata       &   \nodata      &   \nodata      &   O6-7  & -4.13 & -0.21	 &  2011-04-17T08:52:41.0839, SofI \\
 DBS100 Obj4  &    \nodata       &    \nodata       &    \nodata       &   \nodata      &   \nodata      &   G2-3 & -1.18 & 0.35	 &  2011-04-17T08:52:41.0839, SofI \\
 DBS130 Obj1  & 418.99$\pm$25.94 &   0.00: &    \nodata       &   \nodata      &   \nodata      &  \nodata  &  \nodata &  \nodata &  2011-04-18T02:36:44.5824,SofI \\
 DBS130 Obj2  &    \nodata       &    \nodata       &    \nodata       &   \nodata      &   \nodata      &  \nodata &  \nodata  &  \nodata  &  2011-04-18T02:36:44.5824,SofI \\
\enddata
\label{main_par_stars1}
\end{deluxetable}

{ The projected radius of the clusters in our sample is non homogeneously determined in the literature. Based on the much deeper VVV color magnitude diagrams with respect to the previous studies, we determined the visual radius of the clusters performing direct star counting in the $K_S$ band with 10\arcsecspace radius, assuming spherical symmetry. That number is then divided by the area of the rings to determine the stellar density. The projected star number density as a function of radius is shown in Fig.~\ref{radius_all}. The cluster boundary was determined by fitting the Elson et al. (1987) theoretical profile, which represents well the young  clusters. The obtained values are tabulated in  Table~\ref{main_par}, where the errors are errors from the fit.

\begin{figure*}
\epsscale{0.8}
\plotone{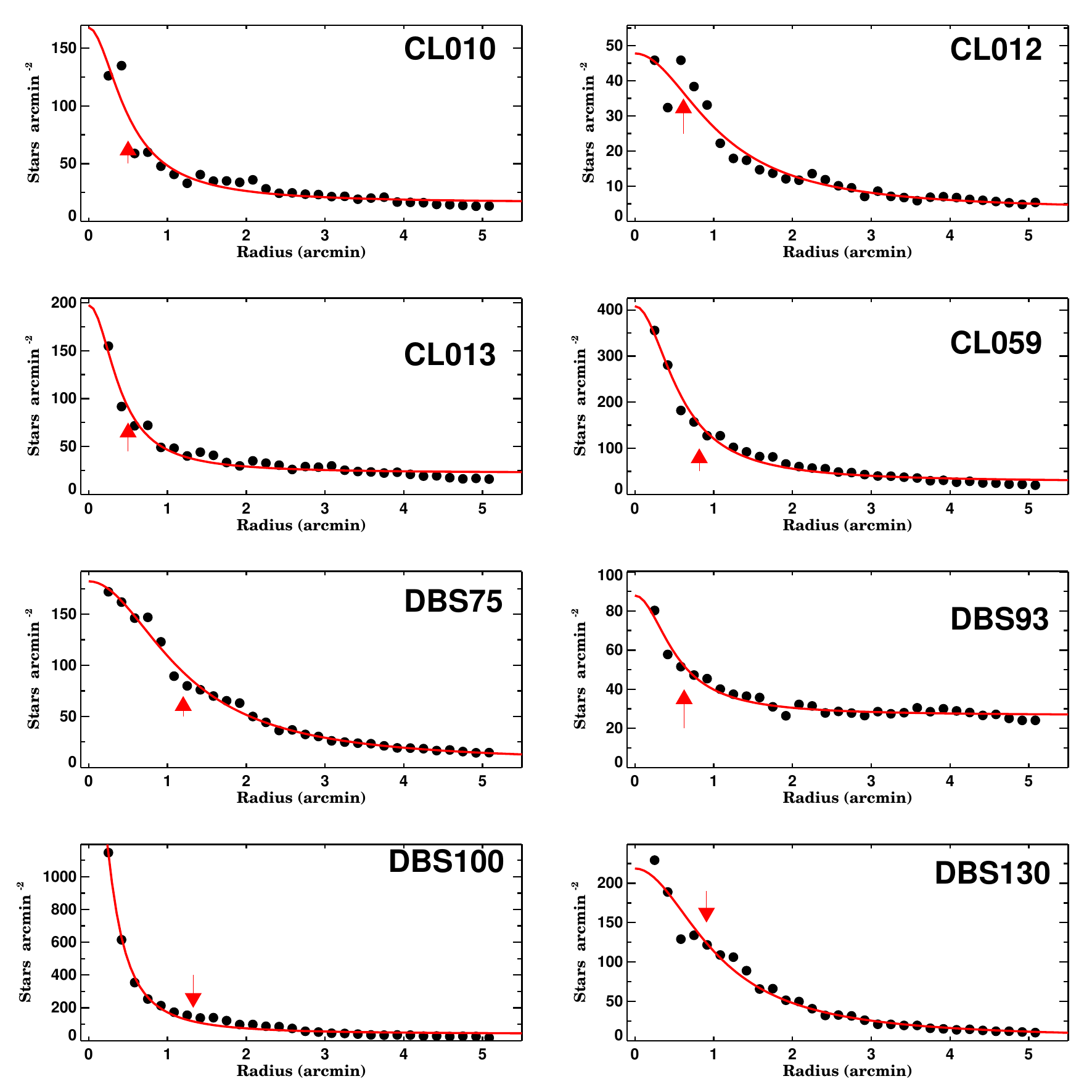}
\vspace{0.1cm}
\caption{The radial density profiles as a function of radius of the clusters in our sample. The solid line stands for the best fit, the arrow marks the obtained radius of the cluster. } 
\label{radius_all}
\end{figure*}

}
\subsection{VVV\,CL010}

\begin{figure}
\epsscale{0.8}
\plotone{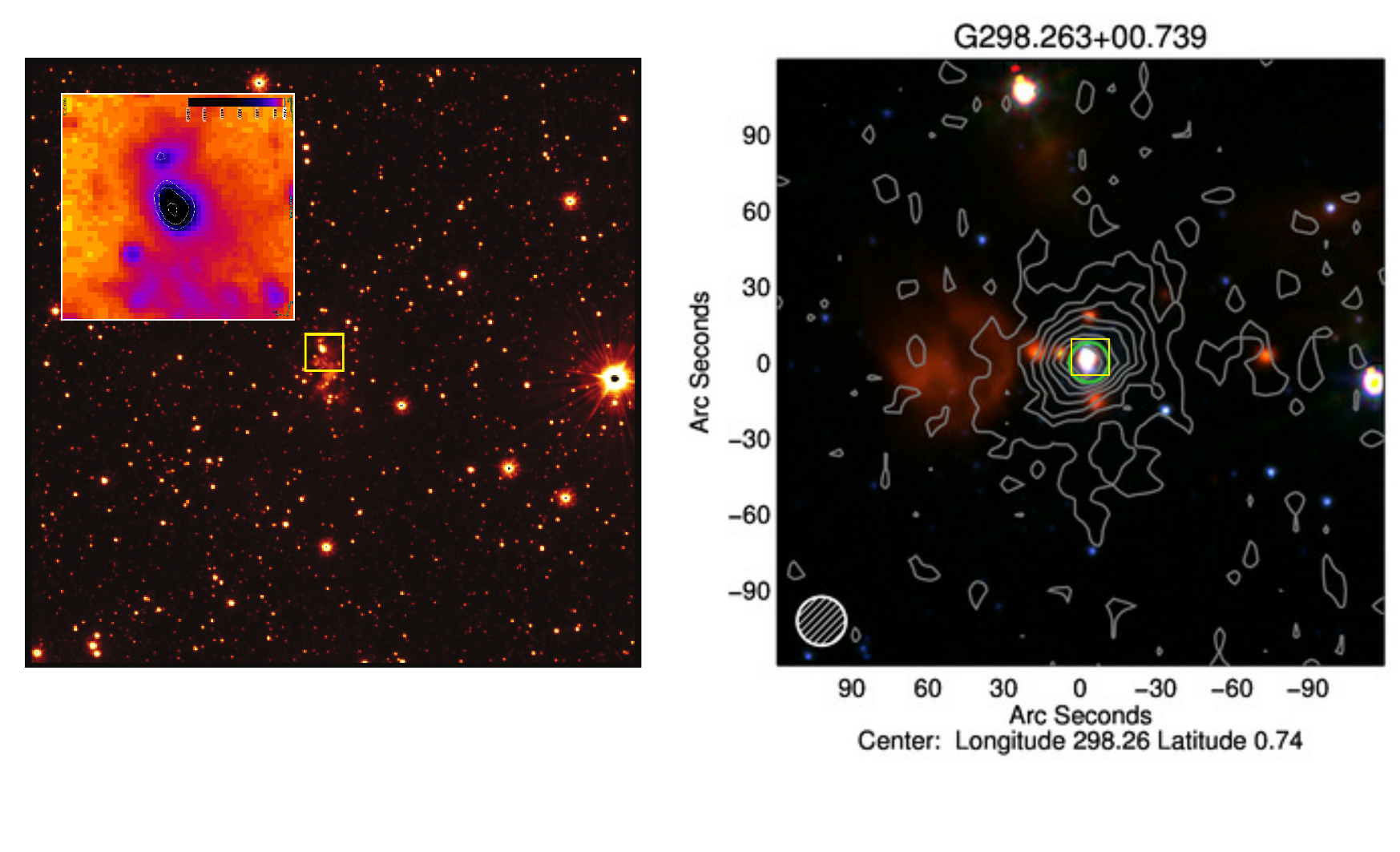}
\epsscale{0.2}
\vspace{0.1cm}
\caption{Ks band composed image of [H SL2000] IRS 1 (Obj\,1).
Right: A GLIMPSE three-color image overlaid with ATLASGAL contours of continumm emission at 870 microns from
Urquhart at al. (2013). } 
\label{Obj1_image}
\end{figure}

VVV\,CL010 was selected from the star-cluster list of Borissova et al. (2011), where it is described as a reddened stellar group, superposed on the strong nebulosity of the H{\sc ii} region GAL 298.26+00.74. The region contains several infrared, maser and mm sources, summarized in the recent paper of Caratti o Garatti et al. (2015) as follows:  the infrared source [HSL2000] IRS 1, coincident with \object{IRAS 12091-6129}, was firstly identified by Henning et al. (2000) at middle infrared (MIR) wavelengths, together with a second source \object{[HSL2000] IRS 2}, located $\sim$28$\arcsec$ westwards. The HII region G298.2622+00.7391 is the dominant source at 8\,$\mu$m. $L_{\rm bol}$ values from the literature range form 1.6 to 5.2$\times$10$^4$\,L$_{\sun}$~ (Walsh et al. 1997; Henning et al. 2000; Lumsden et al. 2013), depending on the adopted distance (3.8--5.8\,kpc). According to these estimates, the source spectral type ranges from B0.5 (Henning et al. 2000) to O8.5~ (Walsh et al. 1997). Both CH$_3$OH (at 6.67 GHz) and OH maser (at 1.665 GHz) emissions are detected towards the source (Walsh et~al. 2001). Very close to [HSL2000] IRS 1 (Cyganowski et al. 2008) observed EGO emission, namely the \object{EGO G298.26+0.74}. Outflow emission from CO (2--1) and CS (2--1) has been reported by Osterloh et al. (1997). The H$_2$ images of Caratti o Garatti et al. (2015), does not show any H$_2$ emission at the EGO position, but a well collimated jet is identified, possibly due to the presence of a multiple system. 
We tried to identify this multiple system on the composed from 49 VVV images (each with a 16 sec exposure time) K$_S$ band image. A plot of [HSL2000] IRS 1 (hereafter Obj\,1) is shown in Fig.~\ref{Obj1_image}. As can be seen, it was not possible to resolve the object, because of the relatively large pixel size (0.339 arcsec) of the VISTA/VIRCAM detector, but the plotted contours clearly show a non stellar image profile. This may be due to a close companion or scattered light. For comparison, on the same plot we show a GLIMPSE three-color image overlaid with ATLASGAL contours of continumm emission at 870 microns from Urquhart at al. (2013), when similar elongation can be seen.  

\begin{figure}
\epsscale{0.9}
\plotone{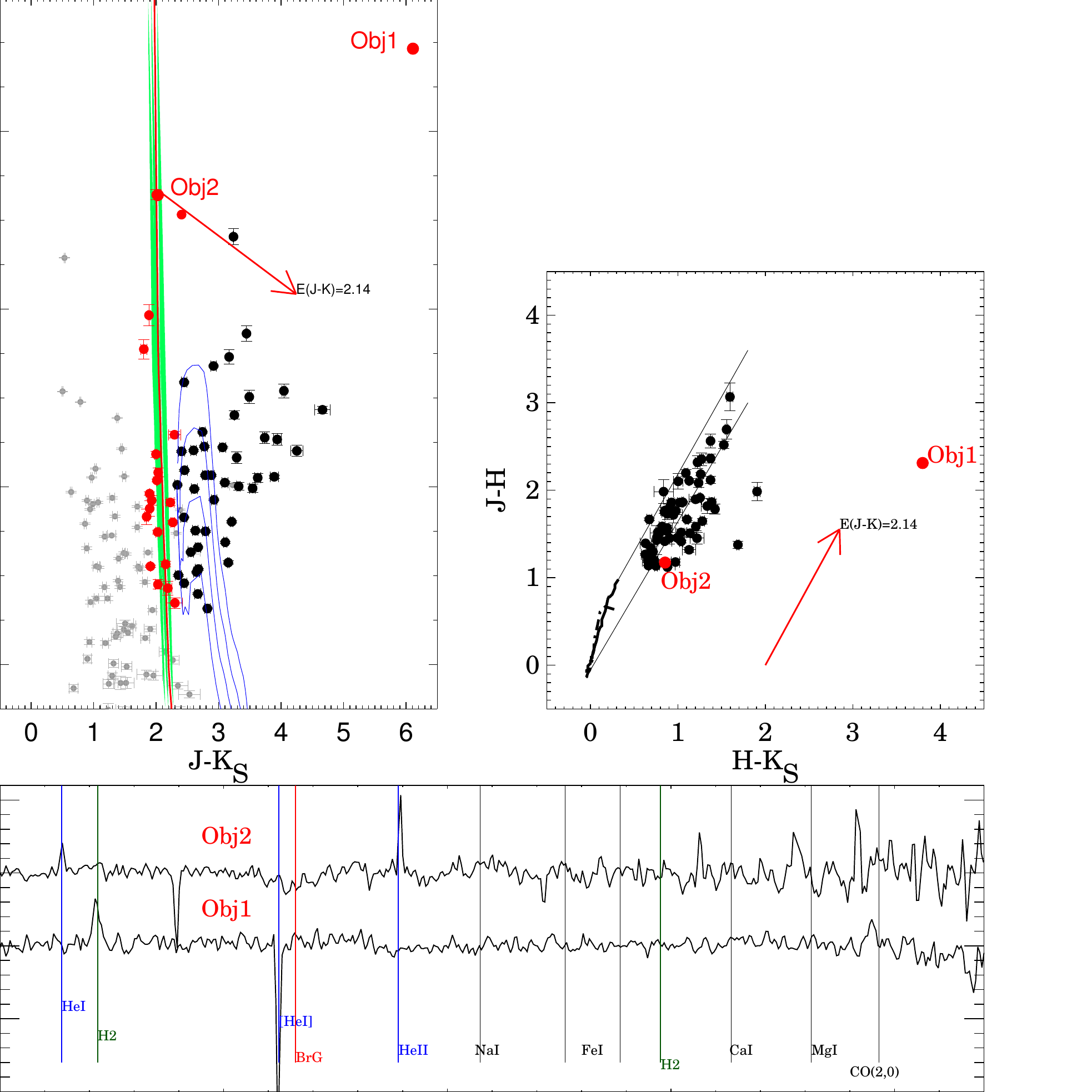}
\vspace{0.2cm}
\caption{Top-left: $(J-K_{\rm S})$ vs. $K_{\rm S}$ color-magnitude diagram for VVV\,CL010. Gray circles are comparison field stars, red and black circles are the most probable main-sequence, PMS, and infrared excess cluster members, after statistical decontamination. Stars with spectra are denoted by red circles and are labeled. The best fit of 2.7 Myr (z=0.020) Geneva isochrone is a solid red line; the green area show the age interval between 1 and 5 Myr; while the blue solid lines stands for pre-main sequence Bell et al. (2014) isochrones for 1.0, 2.0, 4.0 and 8.0 Myr respectively. The red arrow show the reddening vector. The top-right panel shows the $(J-H)$ vs. $(H-K_{\rm S})$ color-color diagram with the reddening vector. Bottom: SofI low resolution spectra of Obj\,1 and Obj\,2.}
\label{cl010_cmd}
\end{figure}

We observed Obj\,1, during our SofI 2012 run, together with  2MASSJ12114653-6146070 (hereafter Obj\,2, Fig.~\ref{cl010_cmd}, lower panel). Both objects exhibit emission lines.  In Obj\,1, the CO v=2-0 first overtone bandhead appears in weak emission; we also detected strong H$_2$ emission as in Caratti o Garatti et al. (2015), the HeI lines are in absorption, and the Br$\gamma$ line is not detected. Following Bik et al. (2006) the first-overtone line CO emission most probably is produced by a circumstellar disk. Obj\,2 shows He\,I/N\,III and He\,II in emission and weak Br$\gamma$ in absorption, and tacking into account its position on the CMD can be classified as O6-8e.
The $(J-K_{\rm S})$ vs. $K_{\rm S}$ diagram of the region is shown in Fig.~\ref{cl010_cmd}. The statistical decontaminated most probable cluster members form a poorly populated main-sequence (MS) and pre--main-sequence (PMS) branches. The group of stars between $(J-K_{\rm S})$ of 3.5 and 5 and $K_{\rm S}$ between 13 and 16 mag can be dusty objects along the line of sight or near infrared excess sources, which can be expected for the star forming regions affected by high levels of differential extinction. Unfortunately, only two of the possible infrared excess sources identified from the color - color diagram (G298.2584+00.7406 and G298.2663+00.7354) have GLIMPSE measurements and can be classified as YSO candidates (see Sec.~5).  The spectroscopically calculated values of $E(J-K)=2.24\pm0.13$ and $(M-m)_{0}=15.4\pm0.8$ (12$\pm4$ kpc) of Obj\,2 were adopted as a first guess for establishing the cluster's reddening and distance via isochrone fitting, and improved estimates of these parameters were obtained through iterative isochrone fitting on the $(J-K_{\rm S})$ vs. $K_{\rm S}$ color-magnitude diagram. The Main Sequence (MS) isochrones for solar metallically (nearly vertical in this mass range) were taken from the Geneva library (Lejeune \& Schaerer 2001) and the Pre-Main Sequence (PMS) isochrones are taken from the Pisa models (Bell et al. 2014). Starting with the spectroscopic reddening and distance estimates, isochrones were shifted along the reddening vector from their intrinsic positions until the best agreement with the observations was achieved. The stars with $K_{\rm S}$ band excesses and with large uncertainties  are removed before doing the fit. The iterations were stopped when the parameters did not change. Uncertainties tied to the cluster reddening and distance were calculated by accounting for the errors of the best fit, with quadratically added photometric errors,
and errors of isochrone degeneracy in $K_{\rm S}$ band for very young clusters. The green area plotted in Fig.~\ref{cl010_cmd} (left) shows the isochrones interval from 1 to 5 Myr, which are practically identical. For VVV\,CL010, a reddening and distance modulus of $E(J-K)$=2.14$\pm0.24$ and $(M-m)_{0}$=14.55$\pm1.3$(8.13$\pm4.8$ kpc) and age of 2.7$\pm1.5$ Myr were than adopted. 

This distance is very different from the kinematic distance of 3.8-5.8\,kpc obtained for Obj\,1 (G298.2620+00.7394) from radial velocity measurements of CO lines (Wu et al. 2004). Following Messineo et al. (2014) we determined the Red Clump (RC) position in the 10x10 arcmin field around cluster center and calculated the RC reddening and distance modulus of 1.4 mag and 13.7 (5.56 kpc) respectively, which is close to the kinematic distance of Obj\,1. Two explanations of this discrepancy are possible: since our distance was calculated using only one spectroscopic parallax, nearly vertical MS isochrone and poorly populated PMS branch, this leads to large uncertainty; and another possibility is that Obj\,1 is not a cluster member. More spectroscopic observations are necessary to clarify this point.

\subsection{VVV\,CL012}

\begin{figure}
\epsscale{0.9}
\plotone{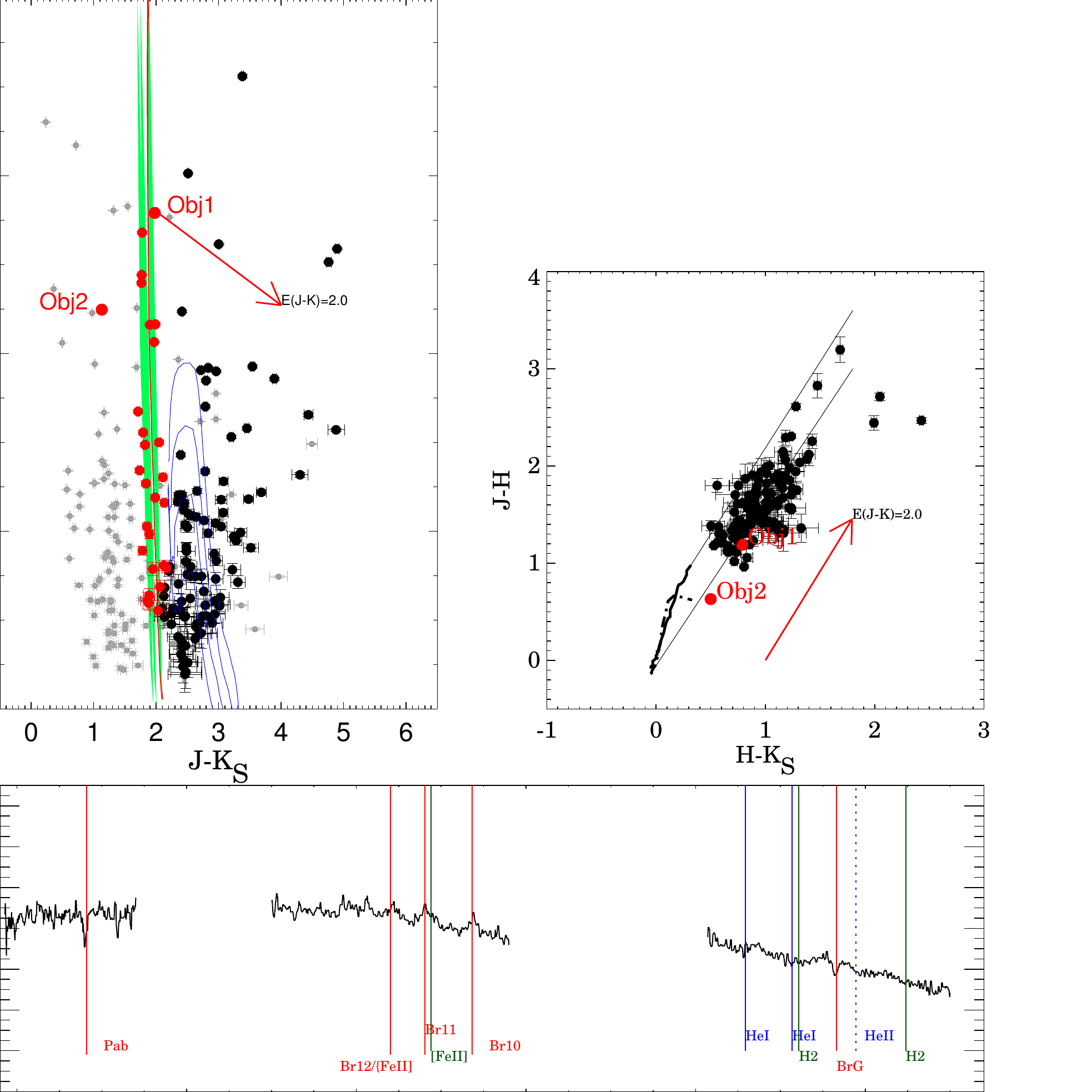}
\vspace{0.2cm}
\caption{Top: $(J-K_{\rm S})$ vs. $K_{\rm S}$ color-magnitude and $(J-H)$ vs. $(H-K_{\rm S})$ color-color diagram for VVV\,CL012. The symbols are the same as in Fig.~\ref{cl010_cmd}. The red solid line shows the best fit of 12 Myr Geneva isochrone.
Bottom: OSIRIS low resolution spectra of Obj\,1.} 
\label{cl012_cmd}
\end{figure}

VVV\,CL012 was selected from the Borissova et al. (2011) list, where it was described as a small, embedded group containing infrared source IRAS 12175-6236. The color magnitude diagram shows stars following Main Sequence and PMS stars and few stars with infrared excess (Fig.~\ref{cl012_cmd}), four of them are measured by Glimpse and satisfy the criterion of Class I/II objects. (see Sec.~5).
The OSIRIS instrument was used to obtain a spectrum of the Obj\,1. As can be seen in Figure~\ref{cl012_cmd}, no He\,II line is identified in the low resolution spectra. The Pa$\beta$ and Br$\gamma$ lines, as well as He\,I  are in absorption, which indicated spectral class not earlier than B2. In the H-band of the spectrum the Brackett series hydrogen lines (H\,I (4-13), (4-12), (4-11), and (4-10)) show weak emission, which can be formed in the surrounding circumstellar material. The EW of the Pa$\beta$ and Br$\gamma$ lines are consistent with B1-2 V spectral type. The combination of spectroscopic parallax values ($E(J-K)=2.1$ and $(M-m)_{0}$=13.55)
with the MS+PMS isochrone fitting yields a reddening and distance modulus for the cluster of $E(J-K)$=2.0$\pm0.3$ and $(M-m)_{0}$=14.1$\pm0.7$ (6.6$\pm2.1$ kpc) and age between 10-12 Myr. The RC distance for this field is calculated to be $(M-m)_{0}$=13.96$\pm0.9$ (6.2$\pm2.7$ kpc), which in this case is in good agreement with the spectro-photometric distance estimate of VVV\,CL012.

\subsection{VVV\,CL013}

\begin{figure}
\epsscale{0.9}
\plotone{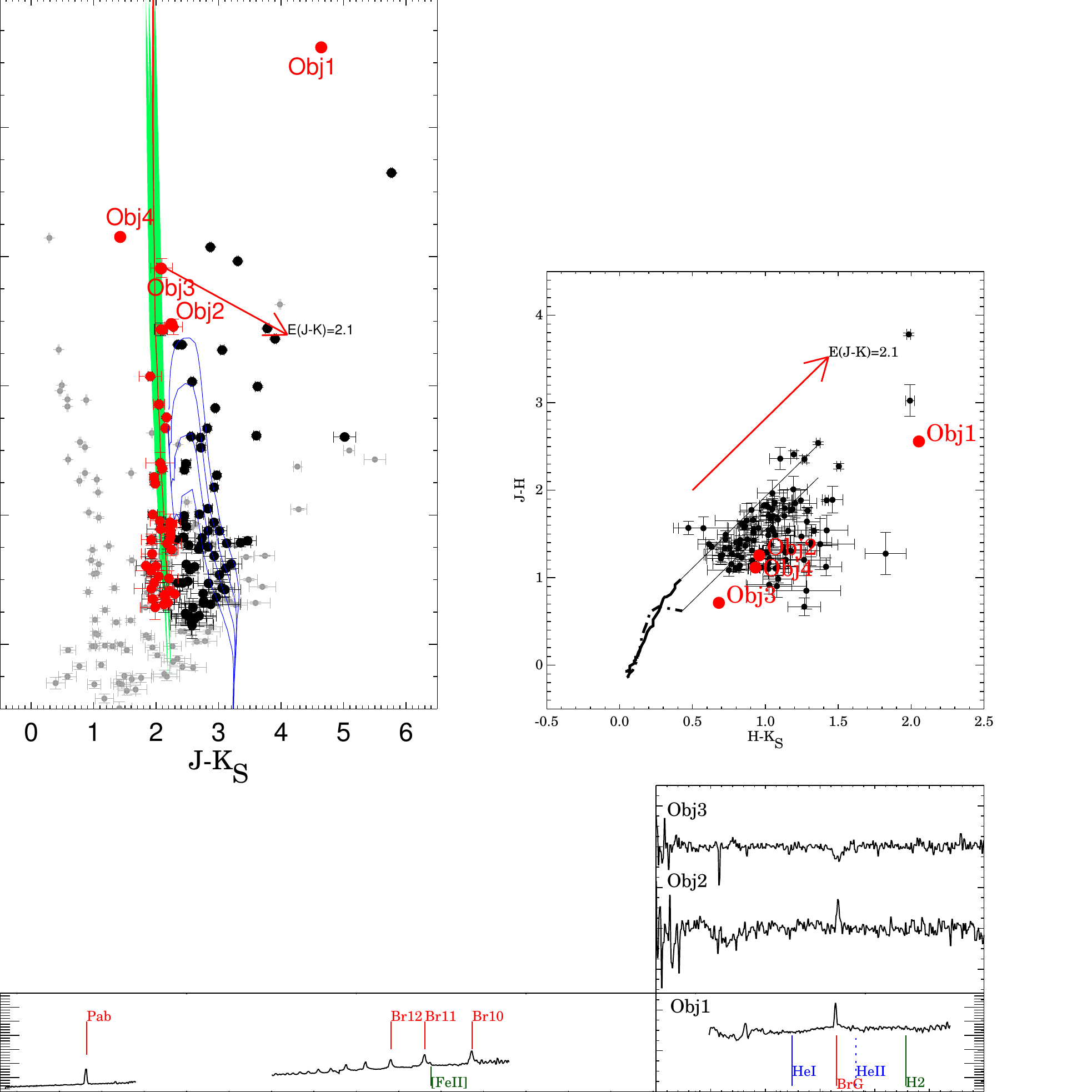}
\vspace{0.2cm}
\caption{The $(J-K_{\rm S})$ vs. $K_{\rm S}$ color-magnitude and and $(J-H)$ vs. $(H-K_{\rm S})$ color-color diagrams for VVV\,CL013. The symbols are the same as in Fig.~\ref{cl010_cmd}, the best fit isochrone is 3 Myr. Bottom: SOAR and SofI low resolution spectra of Obj\,1, \,2 and \,3.} 
\label{cl013_cmd}
\end{figure}

VVV\,CL013 was selected from the Borissova et al. (2011) list, where it is described as a small, embedded cluster, which contains the YSO candidate [MHL2007] G300.3412-00.2190 (hereafter Obj\,1; Mottram et al. 2007). The color magnitude diagram shows a relatively well populated MS and some PMS and infrared-excess stars (Fig.~\ref{cl013_cmd}). The OSIRIS and SofI instruments were used to obtain spectra  of Objs\,1, \,2 and \,3. The spectrum of the Obj\,1, shown in Fig.~\ref{cl013_cmd}, was classified by Mottram et al. (2007) as a young stellar object (YSO) candidate on the basis of 10.4 $\mu$m imaging MID-IR  observations. G300.3412-00.2190 is bright, with $K_{\rm S}$=8.68$\pm$0.03\,mag and very red with $(J-$$K_{\rm S})$=4.61\,mag. Our spectrum shows numerous hydrogen lines in emission of which Pa$\beta$ and Br$\gamma$ (2.17 $\mu$m) are the most prominent. Some He\,I lines can be identified in absorption. These atmospheric spectral features suggest for the central star a spectral type O8-B0\,V. The Obj\,2 shows  only Br$\gamma$ in emission, no other lines are identified in this region, and the object can be classified as a Be star. Obj\,3 shows Br$\gamma$ and He{\sc i} in absorption and is most probably a B2-3\,V star. Following the procedures described earlier, we calculated  the reddening and distance modulus to the cluster as $E(J-K)=2.1\pm$0.3 and $(\rm m-$$\rm M)_0$=13.2$\pm$1.1 (4.4$\pm$2.2 kpc) respectively. The RC distance was not calculated because of very few RC stars in the field. The best fit isochrone gives an age of 2-4 Myr. Additionally, we found in the literature a candidate YSO 2MASS J12201528-6253269 (hereafter Obj\,4; Robitaille 2008). Fifteen stars were selected from our color-color diagram as stars with a possible infrared excess. Nine of them have GLIMPSE measurements and satisfy the criterion for Class I/II objects (Sec.~5).

\subsection{VVV\,CL059}

\begin{figure}
\epsscale{0.9}
\plotone{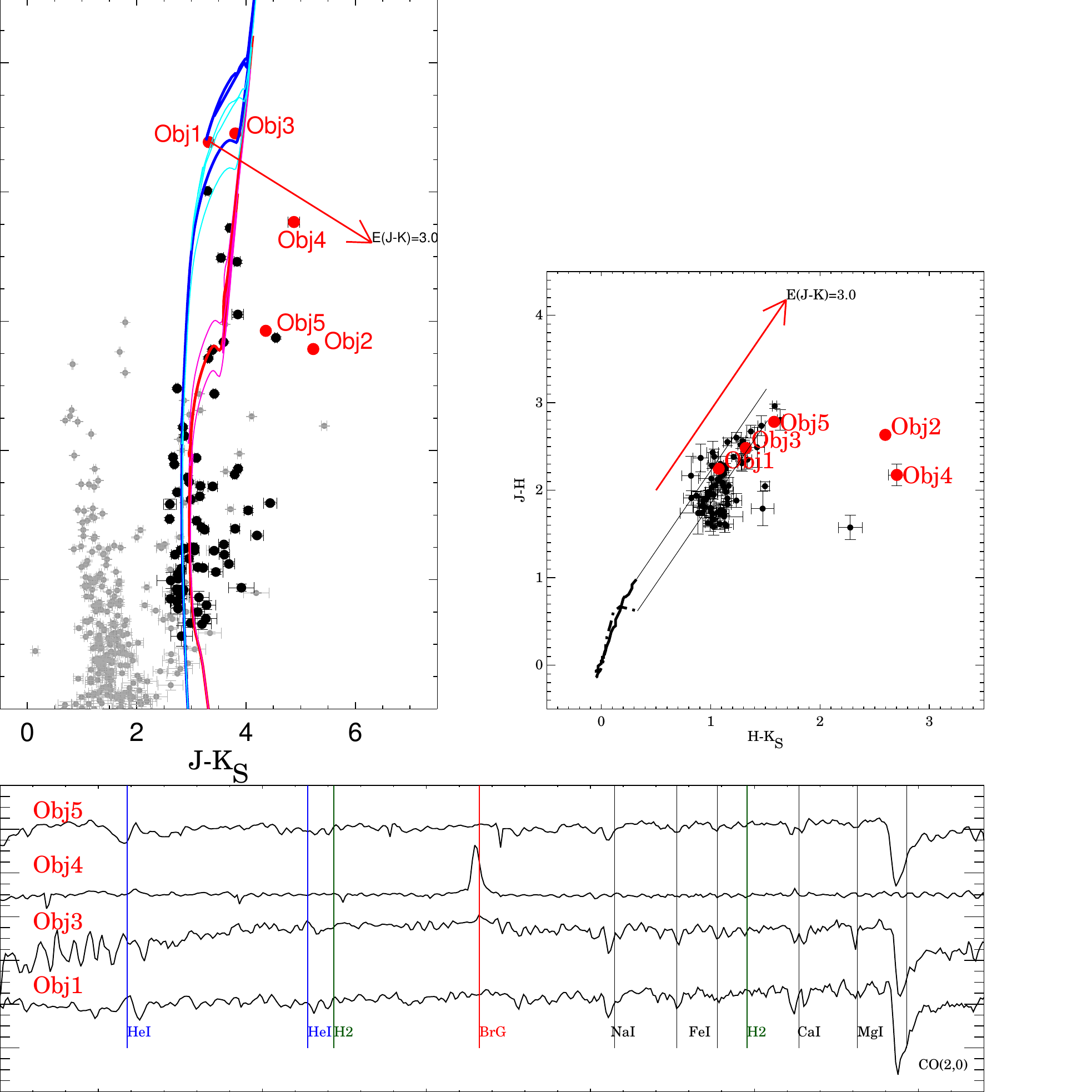}
\vspace{0.2cm}
\caption{The $(J-K_{\rm S})$ vs. $K_{\rm S}$ color-magnitude, $(J-H)$ vs. $(H-K_{\rm S})$ color-color diagrams and ISAAC, VLT medium resolution spectra for VVV\,CL059. The symbols are the same as in Fig.~\ref{cl010_cmd}. The best fits are for 20 Myr (blue) and  316 Myr (red).}
\label{cl059_cmd}
\end{figure}

 VVV\,CL059 was selected from the Borissova et al. (2011) list, where the high reddening of $A_V\approx20$~mag and age between 20 and 30 Myr were determined using only photometry and isochrone fitting. Later on, Morales et al. (2013), pointed out that the cluster must be much younger than 20 Myr and determined a distance of 5.05 kpc, based on comparison with ATLASGAL images. During our ISAAC 2011 run we observed four stars---Obj\,1, 3, 4, and 5---selected from the color-magnitude diagram as possible cluster members. Obj\,1, 3 and 5 show well defined metal lines  of  Ca\,I (2.26$\mu$m), Mg\,I (2.28$\mu$m) and CO-bands, which are similar with a late-K/early-M giant spectral types, and, thus, these stars are classified as  M0\,III, K3-5\,III, and  K0-2\,III, respectively (Fig.~\ref{cl059_cmd}). However, we cannot exclude a luminosity class I classification outright for Obj\, 1 and 3, which is supported by Messineo et al. (2014) Q1 and Q2 indexes. For Obj\,4, on the other hand, Br$\gamma$ shows emission, no H\,I and He\,II lines are detected, and a Ca\,I (2.26 $\mu$m) triplet shows weak emission. Given the lack of helium lines, this Obj\,4 may be a B-star in formation. 
The statistically decontaminated color-magnitude diagram contains 73 possible cluster members and shows two evolved giant/supergiant stars (Obj\,1 and 3), a well defined main sequence, and a couple of stars with infrared excess. Thirteen sources with infrared excess are identified and for 12 of them are identified as YSO candidates (see Sec.~5). Obj\,5 is probably a field star based on its position on the CMD. Additionally, one high amplitude infrared variable from the Contreras Pena (2015) list is found in the field of VVV\,CL059. We calculated  reddening and distance modulus to the cluster as $E(J-K)=3.0\pm$0.2 and $(\rm m-$$\rm M)_0$=13.3$\pm$1.2 mag (4.6$\pm$2.5 kpc), using red giant branch classification of Obj\,1 and 3. The RC distance for the field gives the same value. The best-fit isochrone gives an age of 316$\pm$38 Myr. Respectively, the supergiant classification of these objects, puts the cluster much farther at distance of 11.3$\pm$2.8 kpc, with an age of 20 Myr. According to Morales et al. (2013) CL059 is classified as still being associated with the parent molecular gas, and thus even the age of 20  Myr might be too old for this stage of evolution, considering that stellar feedback could remove the residual gas in a few Myr. Moreover, Obj\,1 has diffuse warm dust/PAH emission in GLIMPSE and is associated with ATLASGAL cold dust emission, which is typical of YSOs.  Our low resolution spectra, however clearly shows metal lines in absorption typical for the evolved stars. Unfortunately, there is not sufficient data (e.g., radial velocities, proper motions, high resolution spectra) to verify cluster memberships of the YSOs, to clarify the nature of Objs\,1 and 3 and reveal the nature of this unusual cluster. 

\subsection{[DBS2003]\,75}

\begin{figure}
\epsscale{0.9}
\plotone{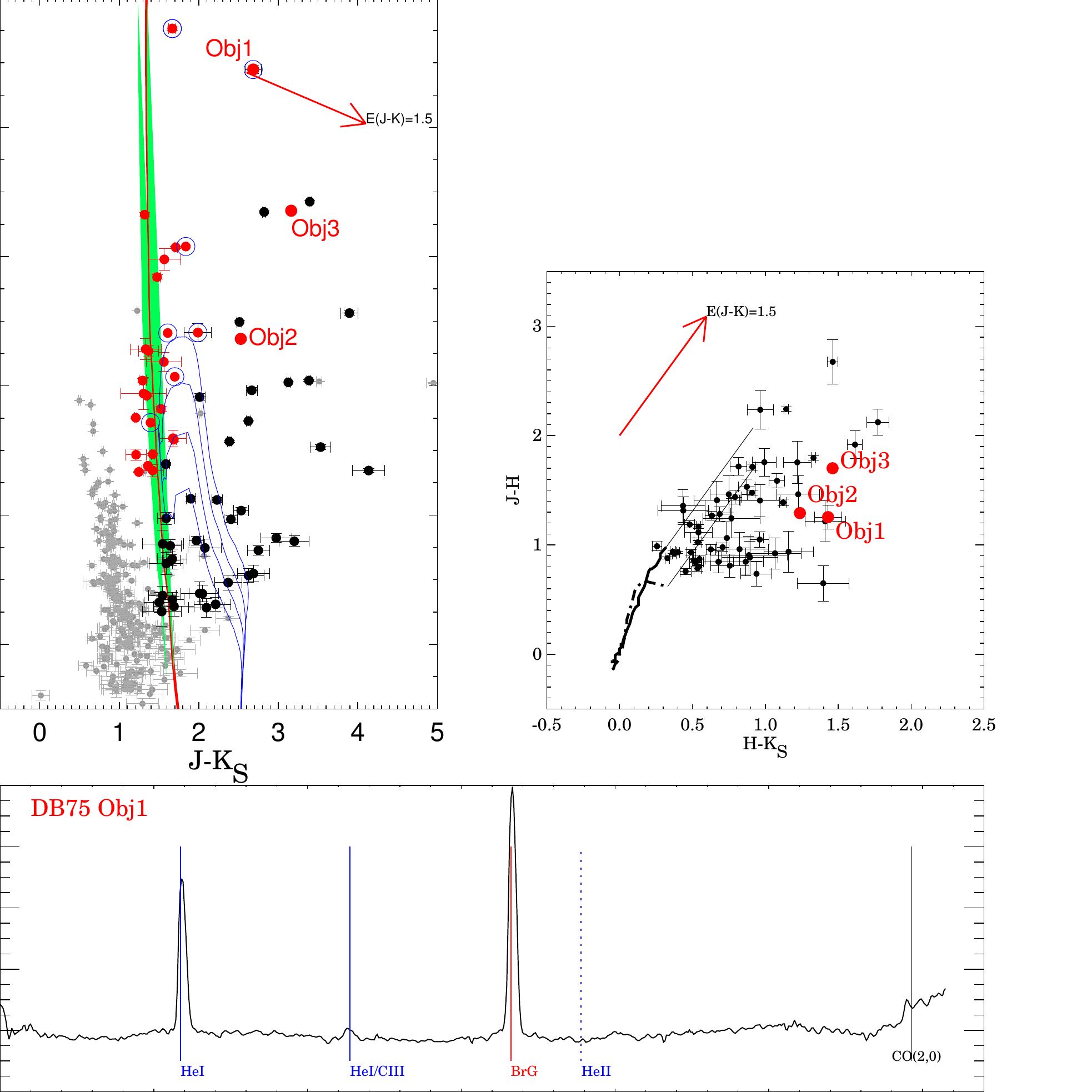}
\vspace{0.2cm}
\caption{The $(J-K_{\rm S})$ vs. $K_{\rm S}$ color-magnitude diagram, $(J-H)$ vs. $(H-K_{\rm S})$ color-color diagram and SofI low resolution spectra for Obj\,1 (2MASS J12090127-6315597) of [DBS2003]\,75. The symbols are the same as in Fig.~\ref{cl010_cmd}, the best fit isochrone is 2 Myr.}
\label{db75_cmd}
\end{figure}

The [DBS2003]\,75 cluster was selected from the Dutra et al. (2003) catalog and is associated with the ESO\,95-1 star-forming region and the ultra-compact H{\sc ii} region IRAS\,12063-6256. The region was first classified as a possible planetary nebulae (Henize 1967), but latter Cohen \& Barlow (1980) suggested that it is an H{\sc ii} region. Obj~1 (2MASS J12090127-6315597) was observed during our SofI 2011 run, and shows strong Br$\gamma$ and H\,I 2.06$\mu$m emissions, but weak emission in He\,I 2.12$\mu$m. CO could be also in weak emission, but it is hard to say because this feature is at the end of the spectral range (Fig.~\ref{db75_cmd}). Our low resolution spectra do not allow us to determine the origin of these emission lines. Thus, it is possible that Br$\gamma$ and H\,I 2.06$\mu$m  are arising in the surrounding HII region, which is also supported by the relatively flat continuum. As pointed out by Cooper et al. (2013) the  H{\sc ii} regions have relatively flat continua, strong  H\,I emission, produced in an optically thin ionized region and, often,  He\,I emission. 
If these emission lines arises from the HII region, than the only visible photosphere line from the YSO will be the weak emission in He\,I 2.12$\mu$m, which indicate early O7-B0 spectral type.

Two stars in the field are classified as YSO candidates in the literature: 2MASS J12090156-6315429 and [MHL2007] G298.1829-00.7860 (Mottram et al. 2007, hereafter Obj\,2 and 3). The statistically decontaminated color-magnitude diagram (Fig.~\ref{db75_cmd}) contains 71 possible cluster members, nine of which were identified in the Kharchenko et al. (2013) catalog as high probability members. The main sequence is poorly populated (only 21 members), the few PMS stars and stars with IR excess are identified. Thus, dispade of relatively large cluster radius, most probably we have a very young, small stellar group, still embedded in dust and gas, rather than an evolved stellar cluster. 
The kinematic distance to the IRAS\,12063-6256 is calculated as 10.5 kpc (Urquhart at al. 2013), the RC distance to the field is $(\rm m-$$\rm M)_0$=13.88 (5.9 kpc), while the spectroscopic distance using the spectral classification of Obj\,1 gives 1.94$\pm$0.9  kpc. The Kharchenko et al. (2013) calculated a reddening of 1.05 mag and a distance of 4.6 kpc. The best isochrone fit favors reddening and distance modulus of $E(J-K)=1.5\pm$0.1 and $(\rm m-$$\rm M)_0$=13.5$\pm$1.0 (5.0$\pm$2.3 kpc) respectively. Thus, we adopted the distance to the cluster an weighted mean of all measurements as 5.6$\pm$3.0 kpc. The stellar group is young, with an age around 2 Myr.

\subsection{[DBS2003]\,93}
\begin{figure}
\epsscale{0.8}
\plotone{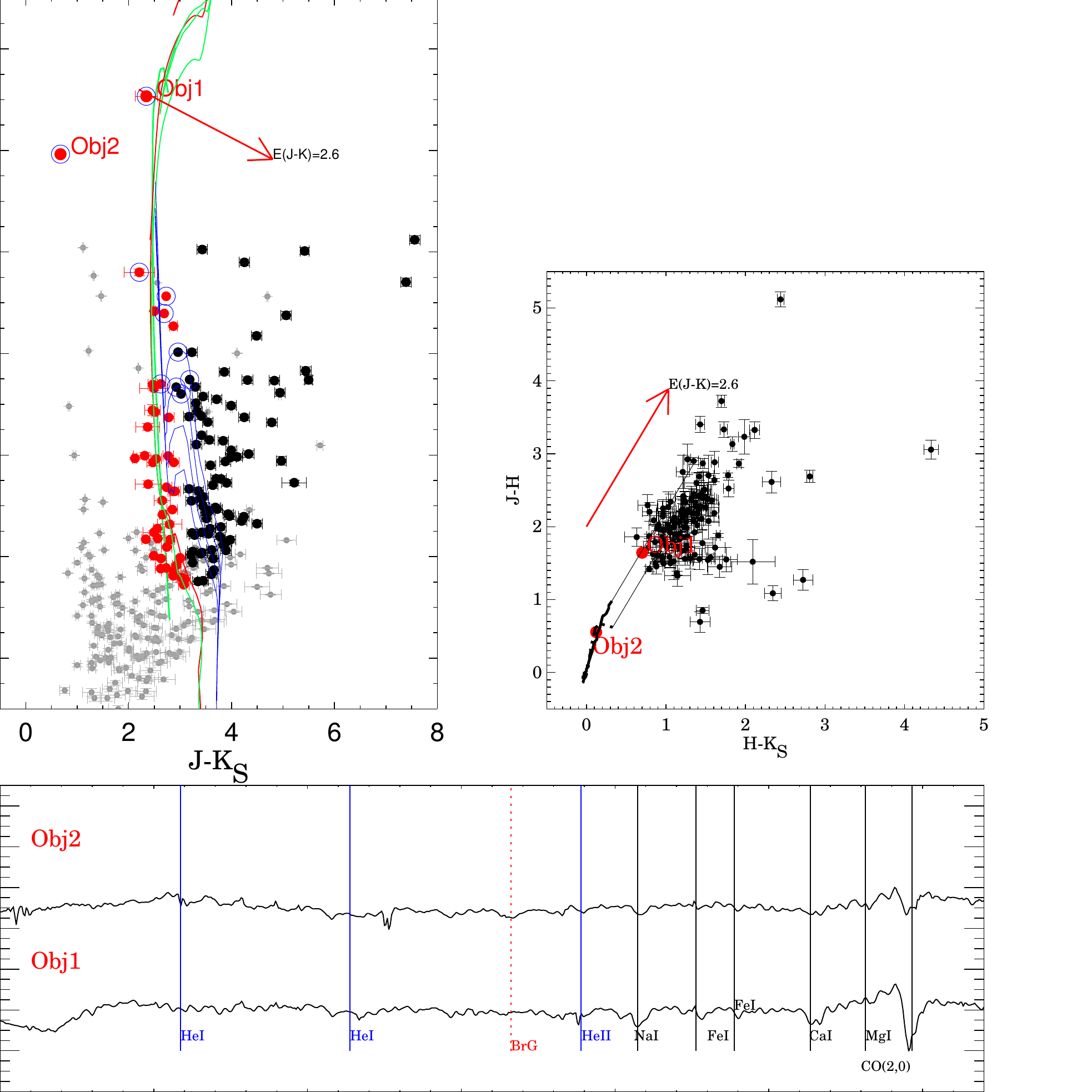}
\vspace{0.2cm}
\caption{The $(J-K_{\rm S})$ vs. $K_{\rm S}$ color-magnitude diagram, $(J-H)$ vs. $(H-K_{\rm S})$ color-color diagram and SofI low resolution spectra for [DBS2003]\,93. The symbols are the same as in Fig.~\ref{cl010_cmd}, blue circles show common stars with Kharchenko et al. (2013) catalog;  the best fit is 20 Myr.}
\label{dbs93_cmd}
\end{figure}

The [DBS2003]\,93 cluster was selected from the Dutra el al. (2003) catalog and is associated with the RCW\,92 star forming region. The color-magnitude diagram shows a couple of red giant branch stars, main sequence stars, and some stars with IR-excess (Fig.~\ref{dbs93_cmd}). Two stars were observed during our SofI, 2011 run, named Obj\,1 and Obj\,2. As can be seen from our low-resolution spectra, Obj\,1 does not show Br$\gamma$; Mg\,I is in weak emission, and the CO shows a inverse P~Cygni profile. The continuum declines toward the red end of the spectrum.  Based on this, we conclude that this is a late M-star in formation. In contrast, Obj\,2 shows shallow and broad Br$\gamma$, the metallic lines are less deep than in Obj\,1 and the CO also has a P~Cygni profile. Thus, the star could be a K-dwarf in formation. Both stars are identified in the Kharchenko et al. (2013) catalog and according to their proper motion analysis are cluster members. Of the cluster members identified by our decontamination, ten MS stars and two YSO candidates (DBS93\,3 and DBS93\,7) are also identified in the Kharchenko et al. (2013) catalog as high probability cluster members (blue circles in Fig.~\ref{dbs93_cmd}). 
Kharchenko et al. (2013) determined a much larger cluster radius, older age and smaller reddening. Based on our 2 magnitudes deeper CMD we determined the visual diameter of the cluster to be 0.72 arcmin. We find the reddening and distance modulus of the cluster to be $E(J-K)=2.6\pm$0.3 and $(\rm m-$$\rm M)_0$=11.62$\pm$0.9 (2.1$\pm$0.87 kpc), respectively. The best fit isochrone gives an age of 20$\pm$0.5 Myr.

\subsection{[DBS2003]\,100}
\begin{figure}
\epsscale{0.8}
\plotone{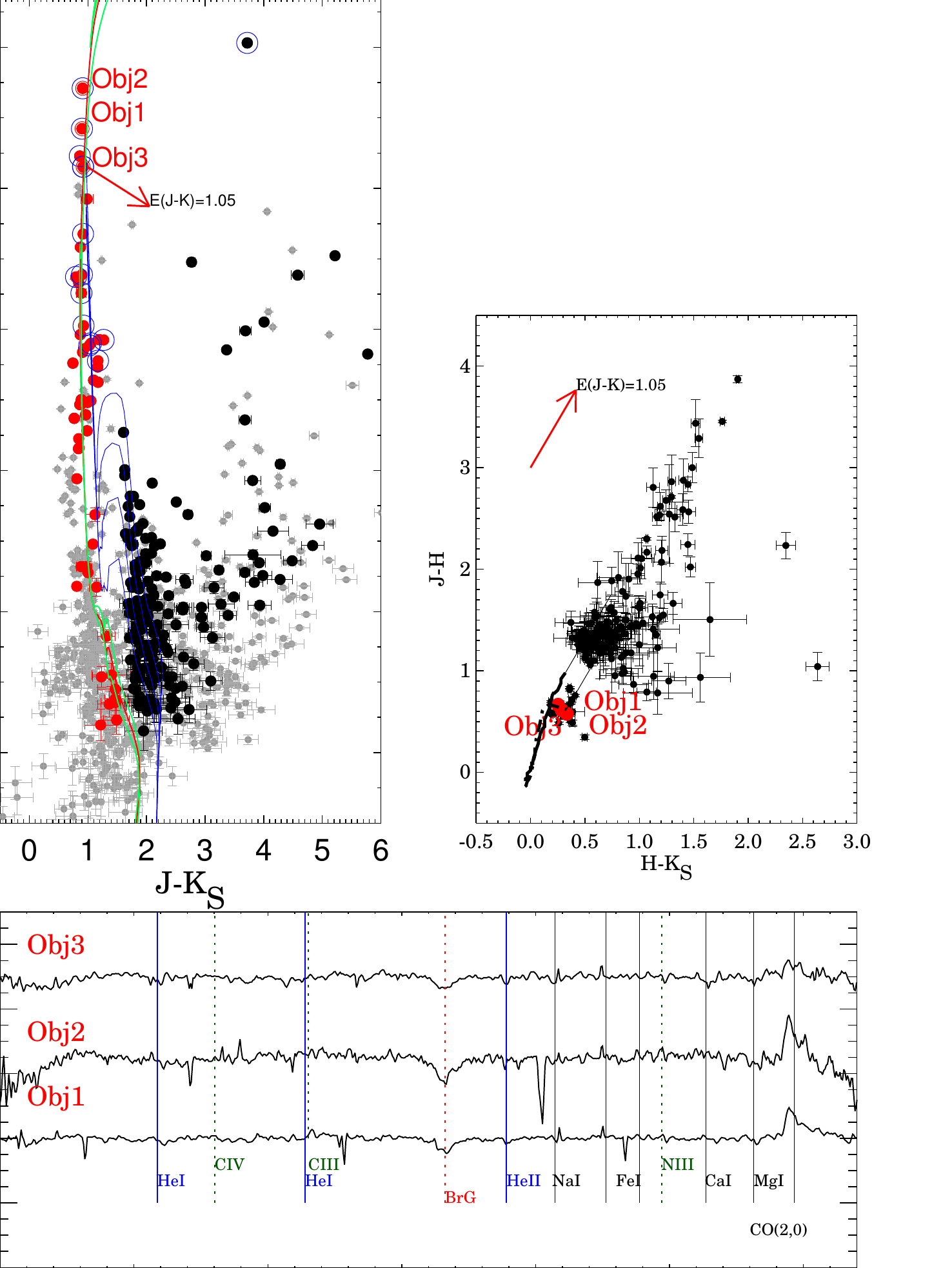}
\vspace{0.2cm}
\caption{The $(J-K_{\rm S})$ vs. $K_{\rm S}$ color-magnitude diagram, $(J-H)$ vs. $(H-K_{\rm S})$ color-color diagram and SofI low resolution spectra for [DBS2003]\,100. The symbols are the same as in Fig.~\ref{cl010_cmd}, the blue circles represent common with Kharchenko et al. (2013) stars; the best fit isochrone is 15 Myr.}
\label{dbs100_cmd}
\end{figure}

The [DBS2003]\,100 cluster was selected from Dutra el al. (2003) and is associated with the RCW\,106 star forming region. The color-magnitude diagram of the cluster is shown in  Fig.~\ref{dbs100_cmd}. It shows a well populated MS and a large number of PMS stars. Some stars with IR-excess can be identified in the color-color diagram (discussed in Sec.~5); two of them DBS100\,ysoc9 and DBS100\,ysoc10 are identified in the Kharchenko et al. (2013) catalog with high membership probability. Three stars (Objs\,1, 2 and 3) were observed with SofI, during 2011 run. All of them show a rather flat continuum shape, Br$\gamma$ He\,I and He\,II in absorption. Objs\,1, 2 and 3 are classified as O4, O6 and O7\,V spectral type, respectively. Despite their very early spectral type all three stars show CO line in emission, which can be associated with a circumstellar disk or envelope. As in the case of DBS\,93 the cluster is part of Milky Way Star Cluster project (Kharchenko et al. 2013), with fundamental parameters as follows: $E(J-K)$=0.52; distance modulus of 2.2 kpc and age of 300 Myr. Our spectroscopically calculated reddening and  distance modulus gives $E(J-K)$=1.1$\pm$0.1 and $(\rm m-$$\rm M)_0$=12.78$\pm$0.8 (3.59$\pm1.3$), respectively. The best fit isochrone confirm the derived spectroscopic distance and reddening and gives an age of 10-15 Myr. As in the case of [DBS2003]\,93, based on our much deeper CMD, we determine this cluster to be much younger, smaller and redder than indicated in the Kharchenko catalog.

\subsection{[DBS2003]\,130} 

\begin{figure}
\epsscale{0.8}
\plotone{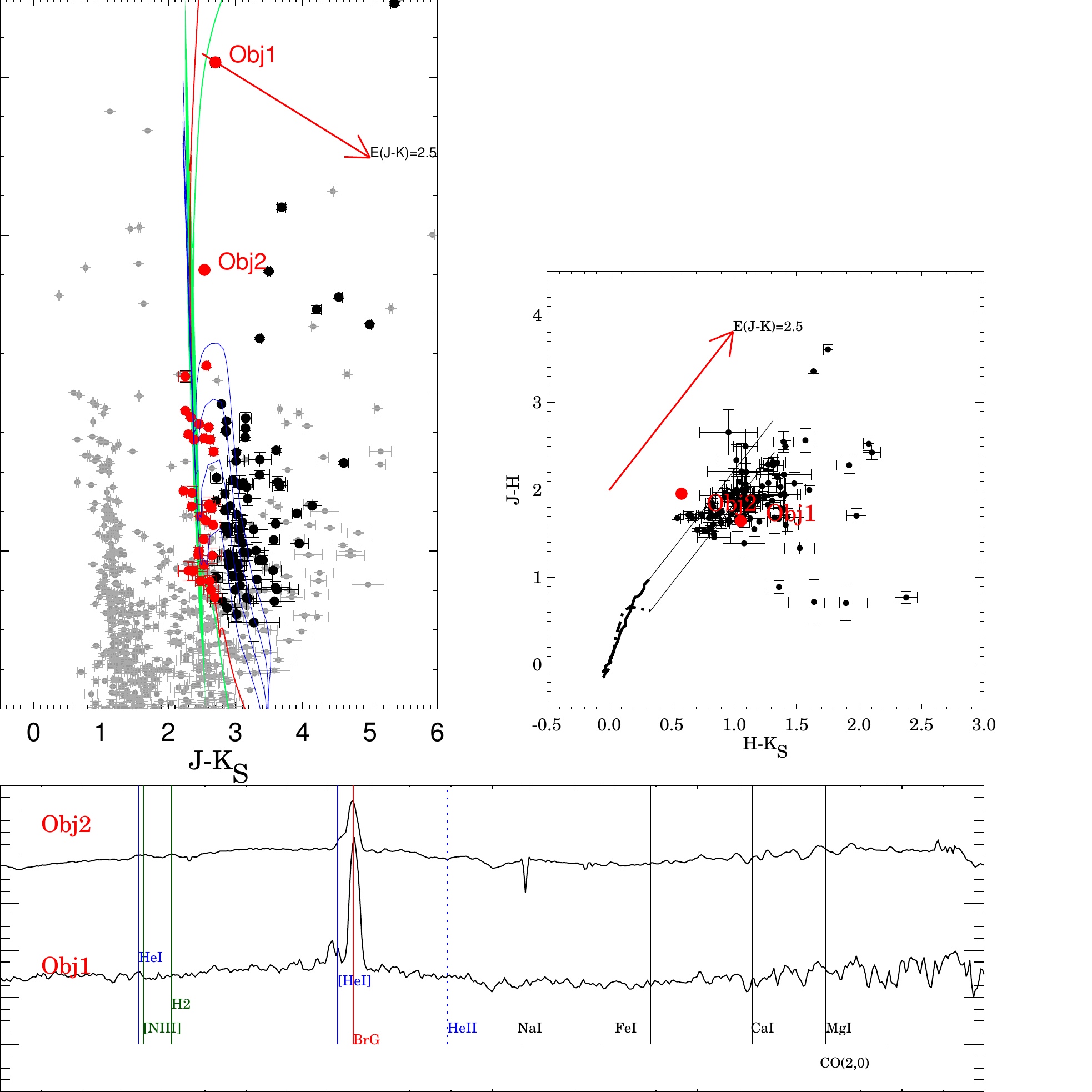}
\vspace{0.2cm}
\caption{The $(J-K_{\rm S})$ vs. $K_{\rm S}$ color-magnitude diagram, $(J-H)$ vs. $(H-K_{\rm S})$ color-color diagram and SofI low resolution spectra for [DBS2003]\,130. The symbols are the same as in Fig.~\ref{cl010_cmd}, the best fit isochrone is 3 Myr.}
\label{dbs130_cmd}
\end{figure}

The [DBS2003]\,130 cluster was selected from the Dutra et al. (2003) catalog and is associated with the G305 star forming region. The color-magnitude diagram of the cluster is shown in Fig.~\ref{dbs130_cmd} and shows a well populated MS, some amount of PMS and stars with IR-excess (see Sec.~5). Two stars (Objs\,1 and 2) were observed with SofI, during 2011 run. As can be seen, Obj\,1 shows Br$\gamma$ and  He\,I (2.06$\mu$m) in strong emission, however, the absence of He\,II would imply an spectral type not earlier than O8. Thus, we assign B0Ie for the spectral type of this star. The spectrum of Obj\,2 is similar, but Br$\gamma$ and  He\,I are less stronger, welding to B0\,Ve spectral type.  The combination of spectral parallax and isochrone fit gives a reddening and distance modulus to the cluster of $E(J-K)=2.5\pm$0.2 and $(\rm m-$$\rm M)_0$=13.1$\pm$1.2 (4.17$\pm$2.3 kpc), which is consistent with G305 region distance. The best fit isochrone gives an age of 3-5 Myr. 

\section{Stellar mass of the clusters}

\begin{figure}
\epsscale{0.3}
\plotone{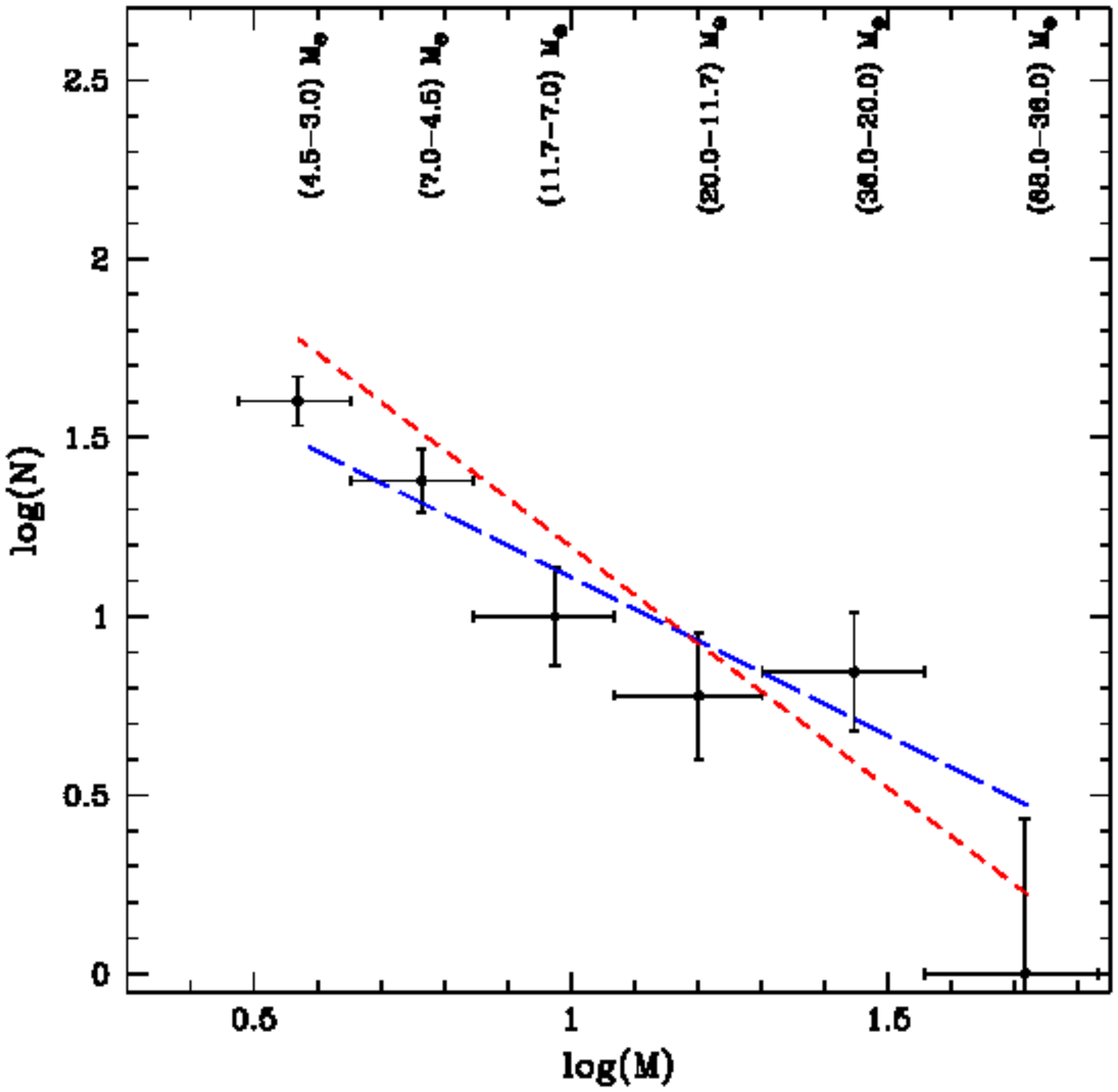}
\plotone{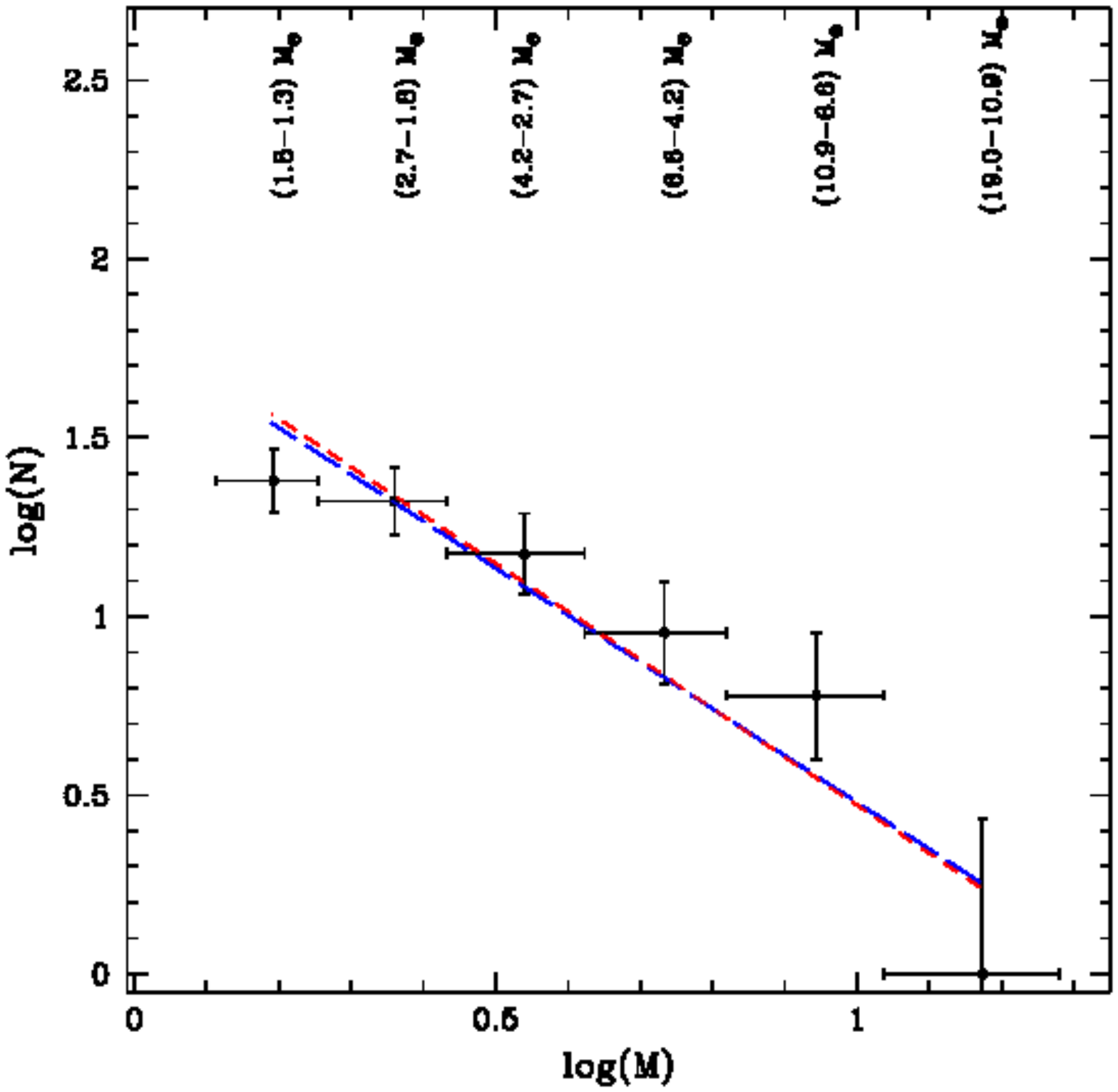}
\plotone{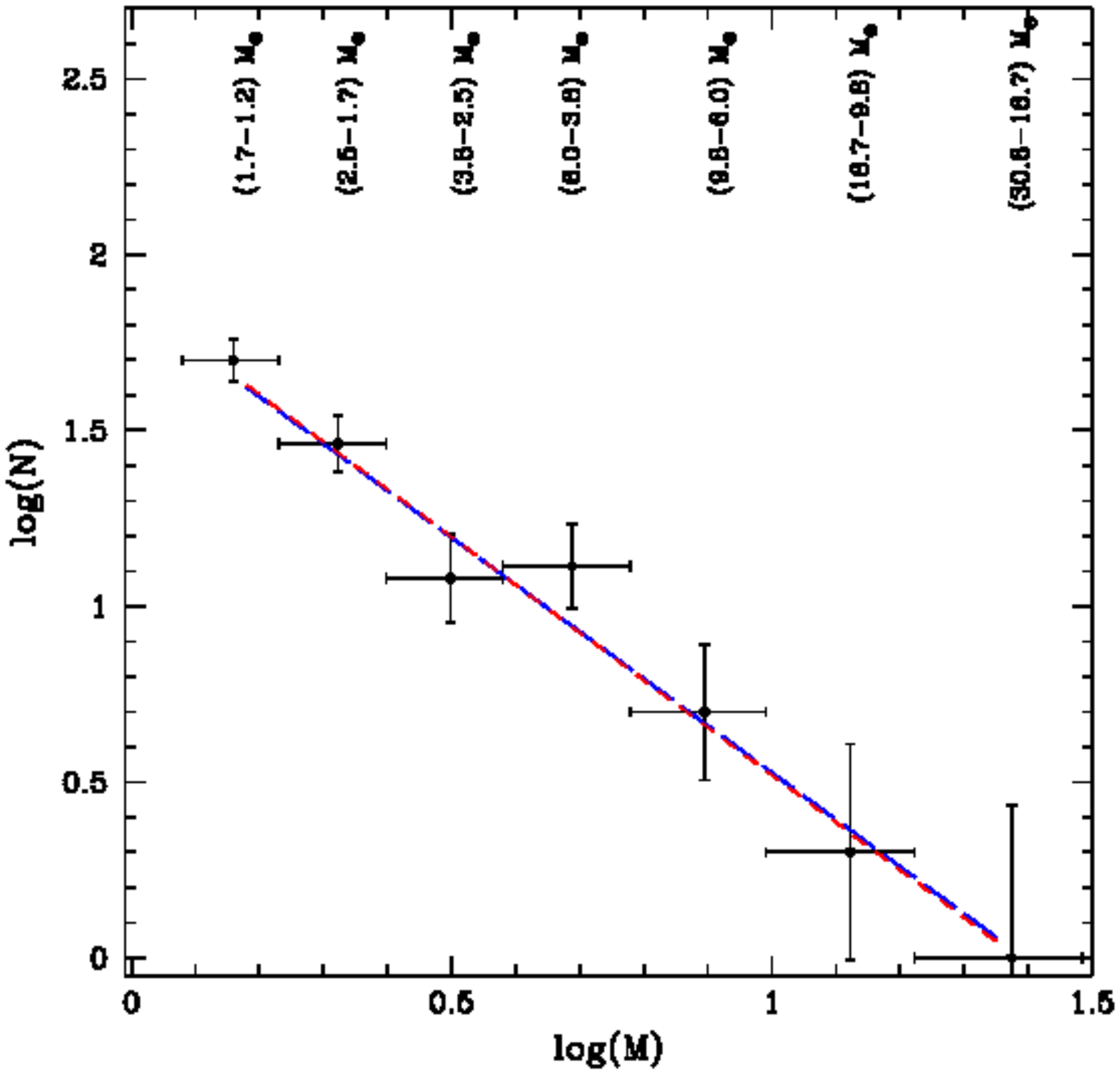}
\plotone{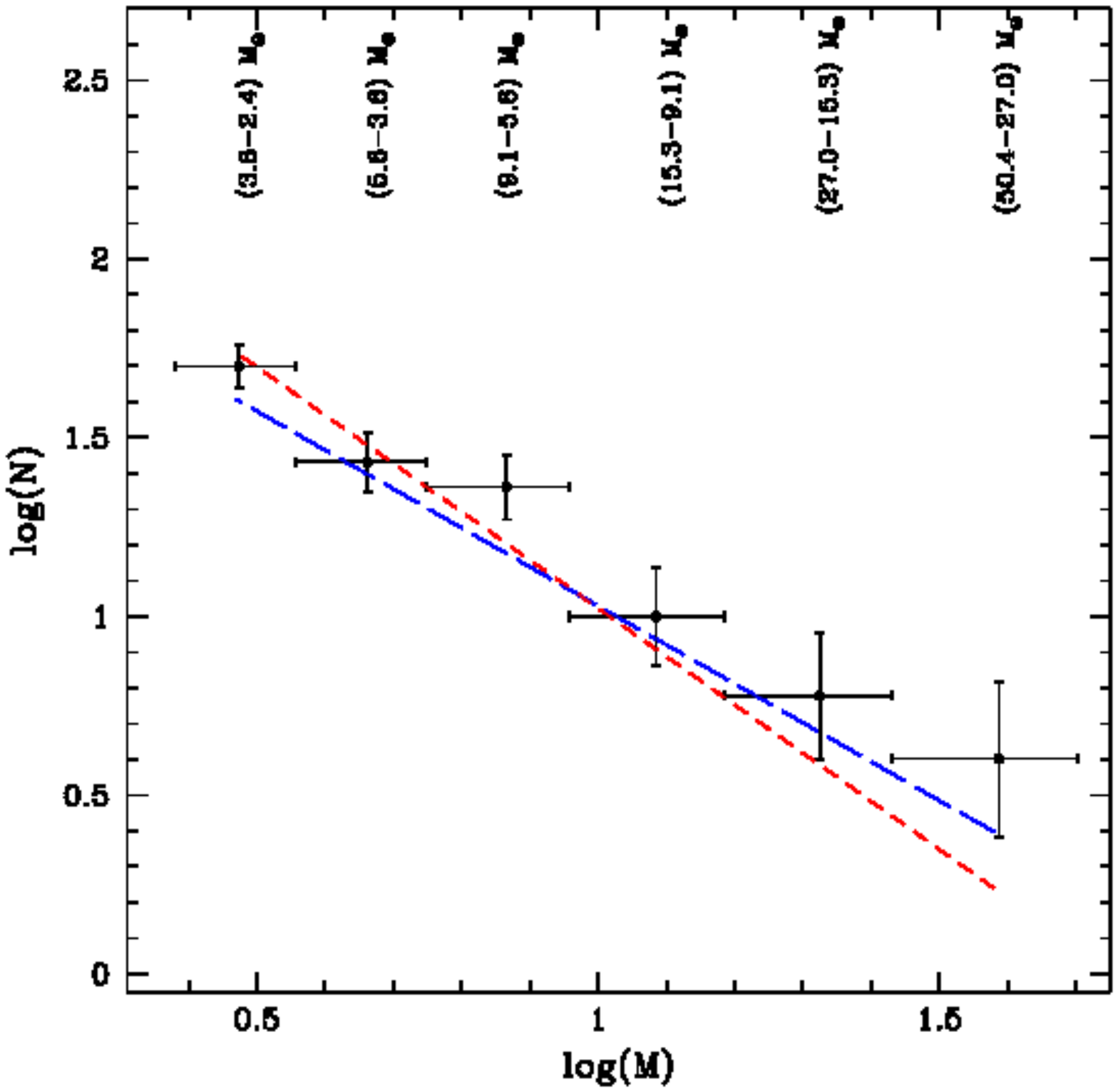}
\plotone{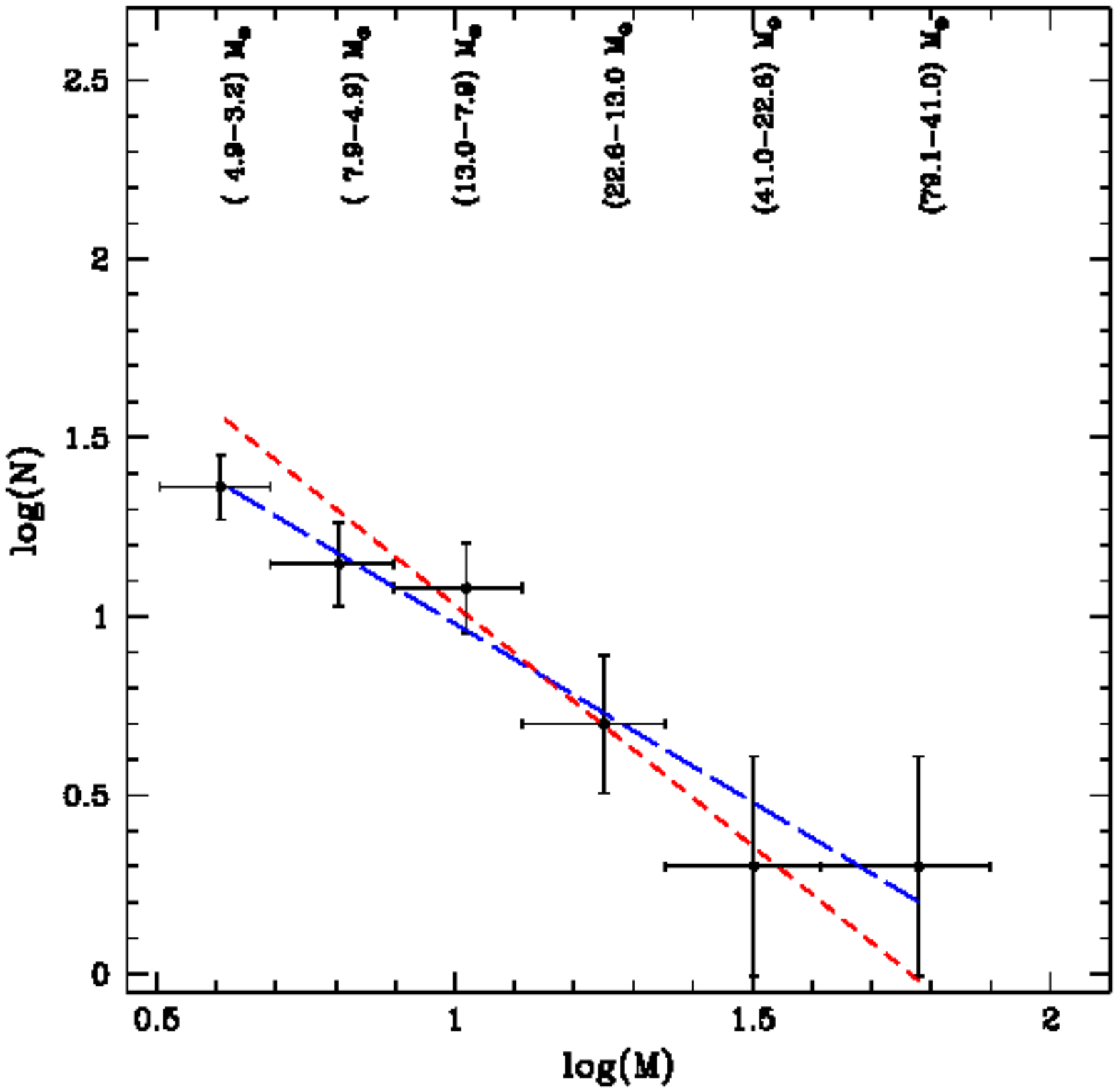}
\plotone{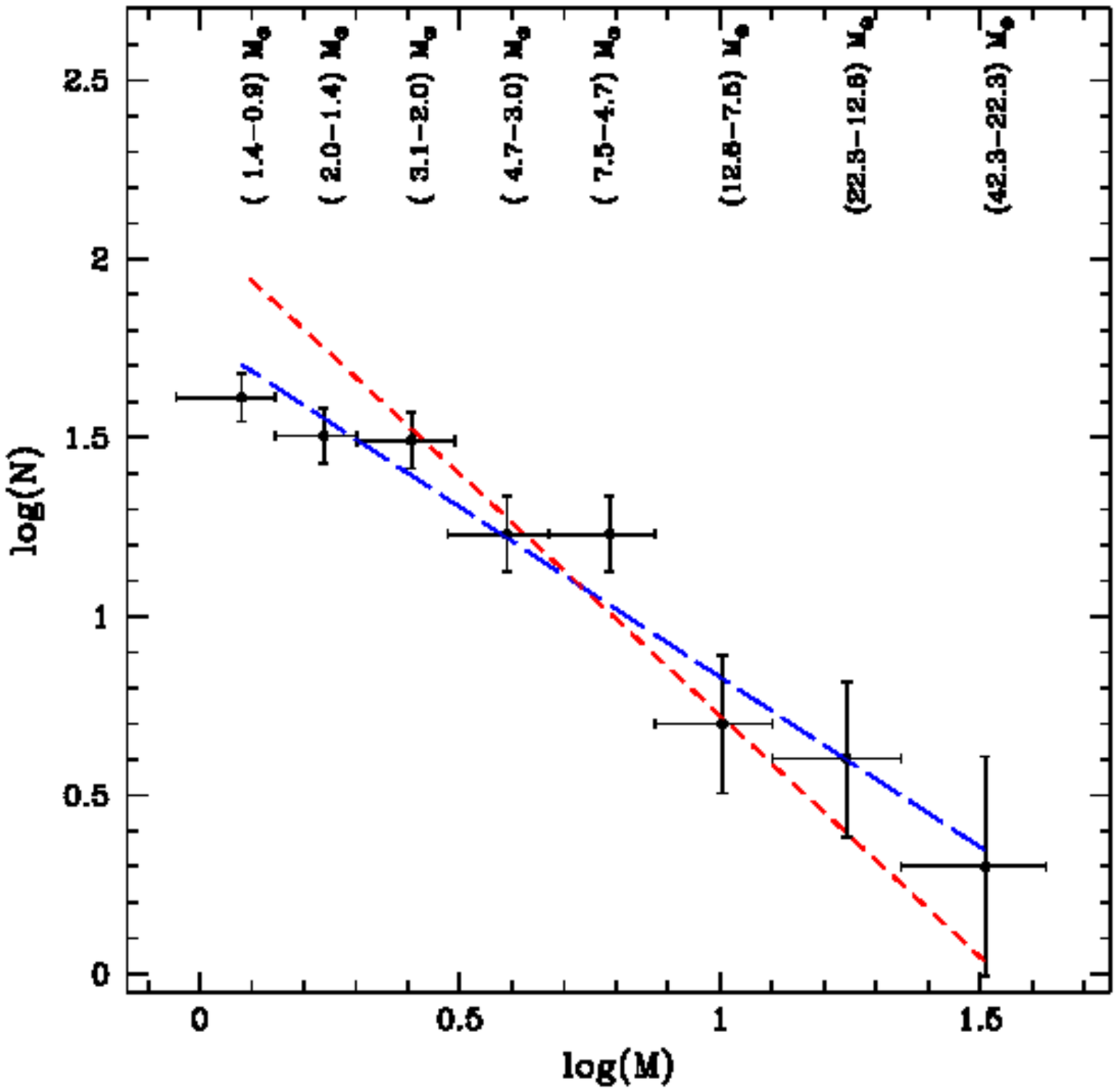}
\plotone{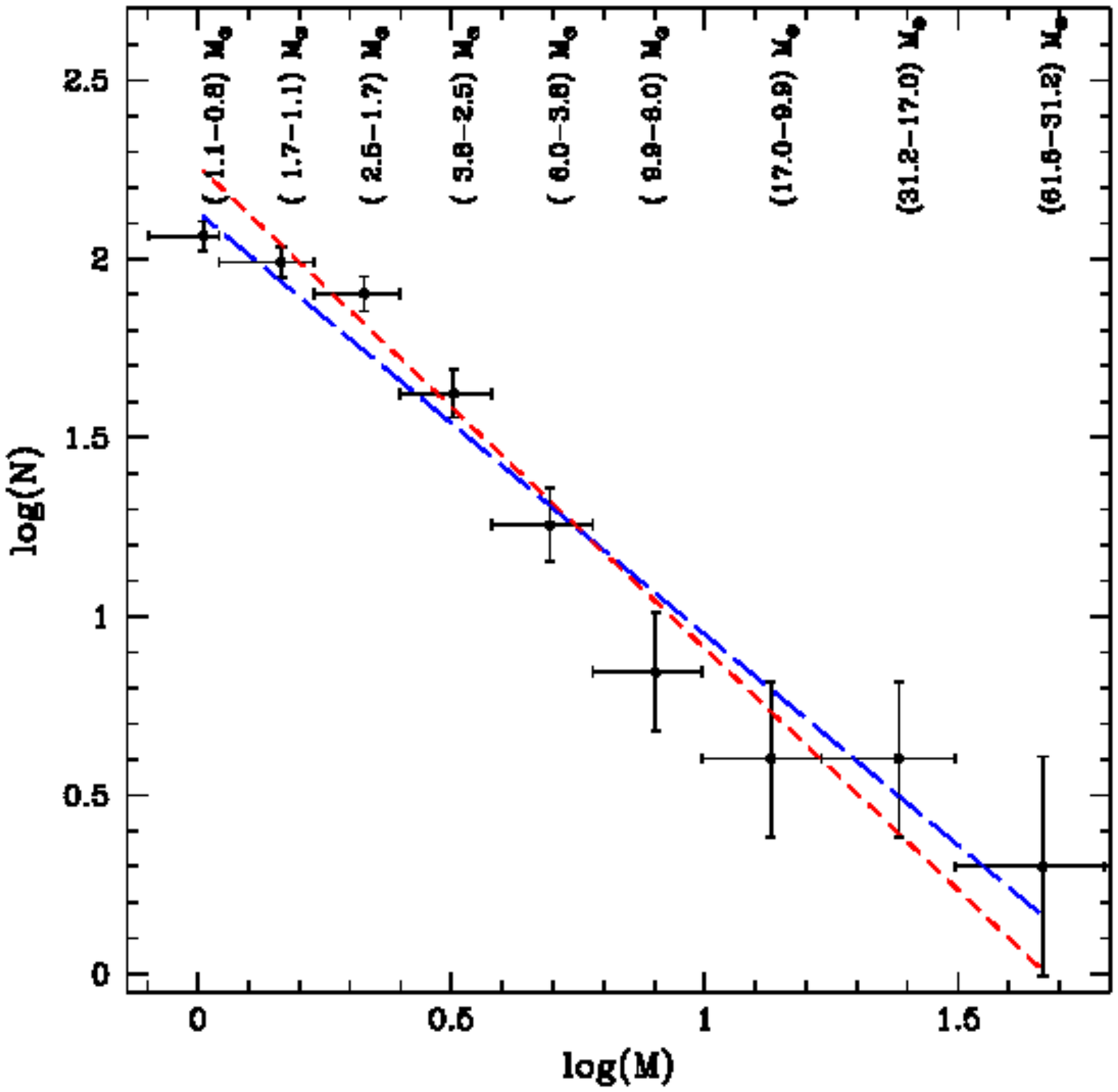}
\epsscale{0.4}
\plotone{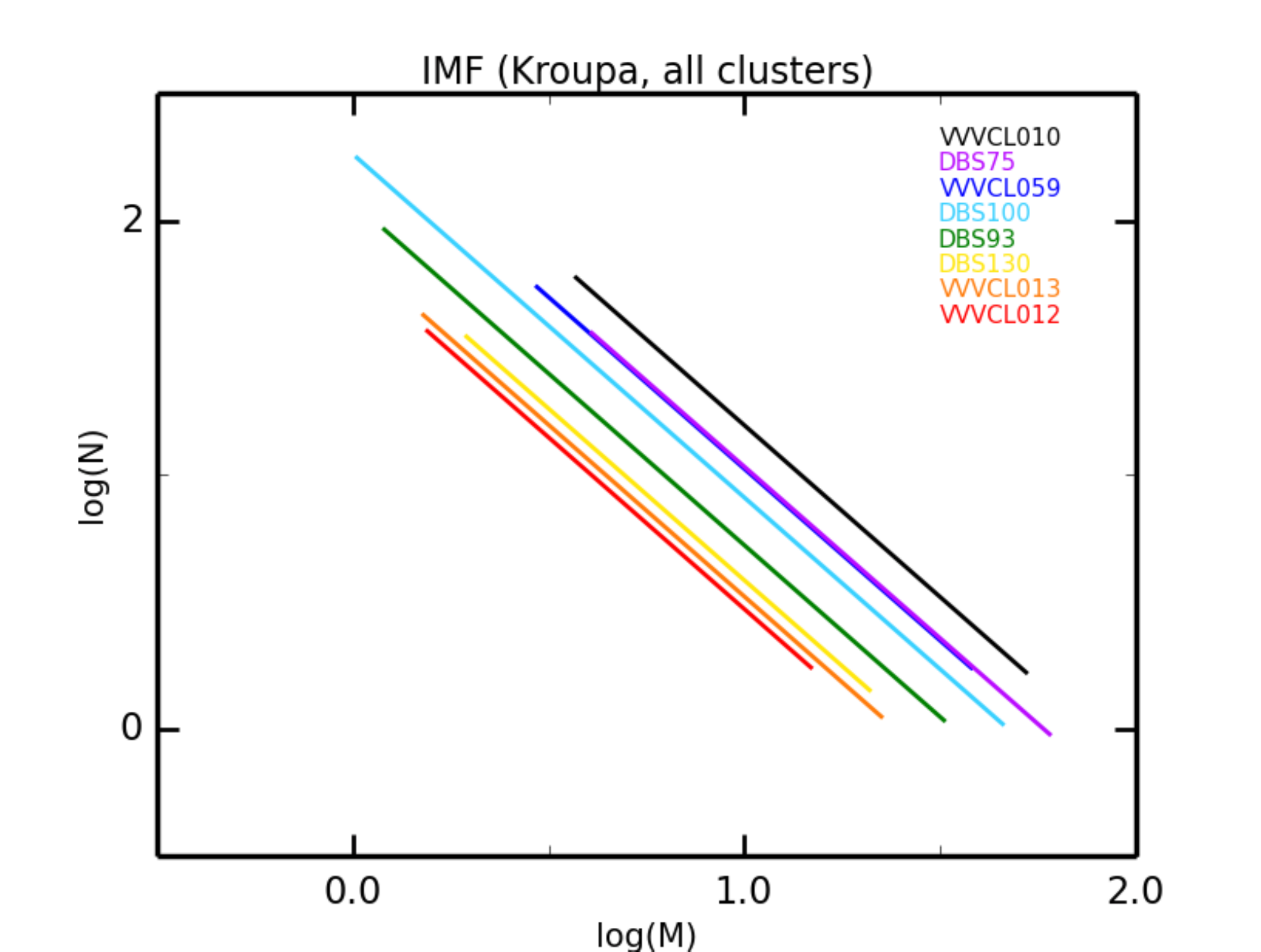}
\vspace{0.2cm}
\caption{The present-day mass function of clusters from our sample: from left to right VVV\,CL010, VVV\,CL012, VVV\,CL013, VVV\,CL059 and [DBS2003]\,75, [DBS2003]\,93 and [DBS2003]\,100.The points show the central position in the mass ranges indicated above them, and the red line corresponds to the Kroupa IMF, while the blue one stands for the best fit of the data. 
Bar sizes indicate the mass bin equivalent to each magnitude bin (from the luminosity function) of 1 mag in $K_{\rm S}$. The last figure gives the summary of the slopes. } 
\label{mf_hist}
\end{figure}

To estimate the total cluster masses, we first constructed the cluster present-day mass function (PDMF) using the CMD and then integrated the initial mass function (IMF) fitted to the cluster PDMF.  We obtained the cluster present-day mass function by projecting the main sequence most probable cluster members, following the reddening vector, to the main sequence located at the corresponding distance. The main sequence is defined by the colours and magnitudes given by Cox (2000). The slope $\Gamma$ of the obtained present-day mass functions of the clusters is given in Table.~\ref{main_par}. As can be seen, the $\Gamma$ values are close to the Kroupa IMF Kroupa (2001).  After deriving the cluster present-day luminosity function, using 1 $K_{\rm S}$-mag bins, we converted the $K_{\rm S}$ magnitudes to solar masses using values from  Martins et al. (2005) for O-type stars and from Cox (2000) for stars later than O9.5\,V. The present-day mass functions, shown in Figure \ref{mf_hist}, are fitted and integrated between 0.1~M$_{\odot}$ and the most massive member candidate in each cluster. The corresponding masses are given in Table.~\ref{main_par}, where the errors corresponds to the fitting of the IMF to the data, and includes also reddening and distance errors. All clusters in the sample are low or intermediate mass, showing masses between 110 and 1800 ${M}_{\odot}$.

\section{Search for YSO candidates, variability and Spectral energy distribution}

\begin{figure*}
\epsscale{0.8}
\plotone{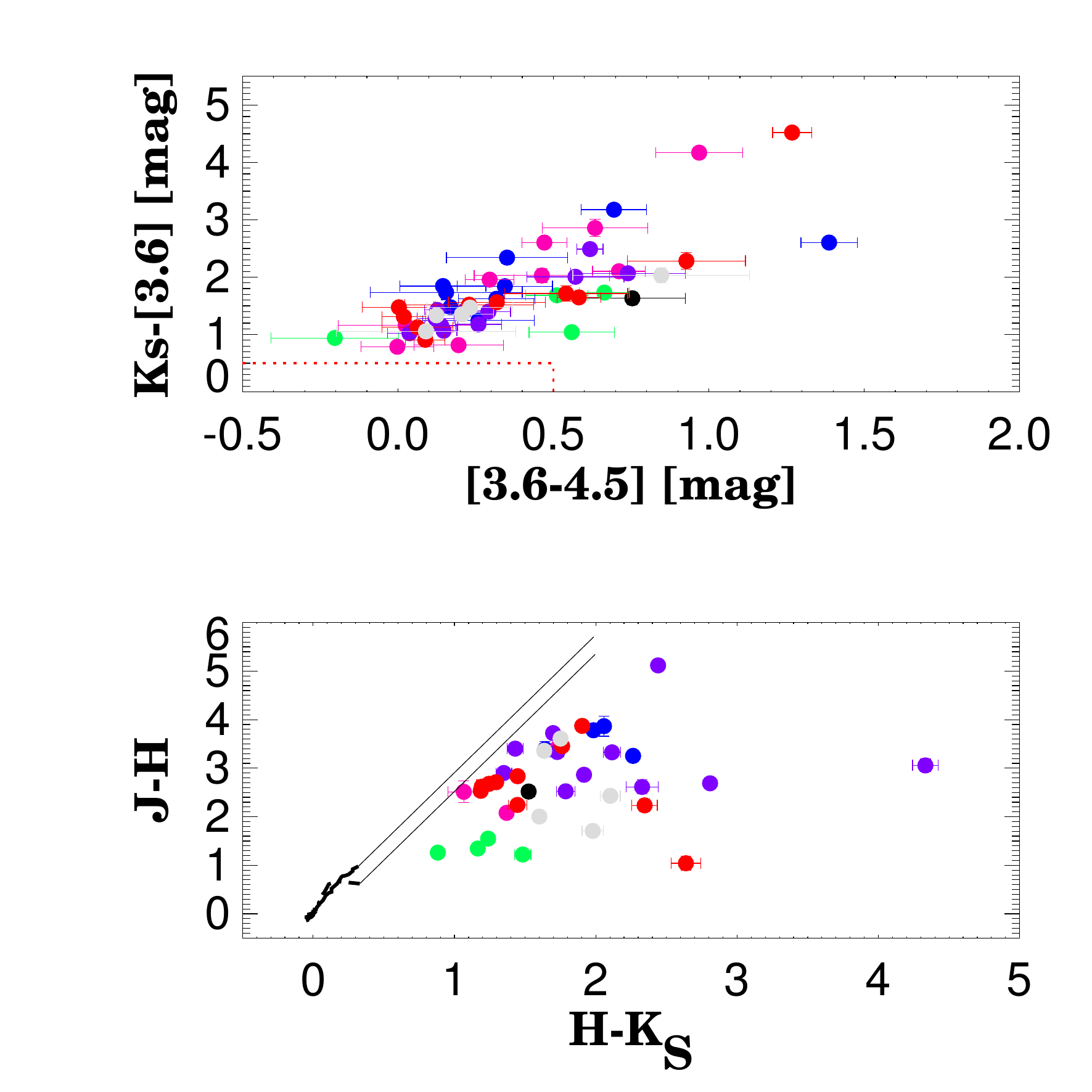}
\vspace{0.1cm}
\caption{Upper panel: The $K_{\rm S} - [3.6]$,$[3.6] - [4.5]$ colour-colour plot of 48 variable stars that are detected in GLIMPSE I. The red dashed lines represent the limits used to select class I and class II YSOs. The YSO in the clusters are color-coded as follow: VVV CL\,010:black; Cl\,012:green; CL\,013:blue; Cl\,059:magenta; DBS\,93:pink ; DBS\,100:red; DBS\,130:gray.
Lower panel: The $(J-H)$ vs. $(H-K_{\rm S})$ color-color diagram of the sample. The continuous and dashed lines represent the sequence of the zero-reddening stars of luminosity classes I (Koornneef 1983) and V (Schmidt-Kaler 1982).} 
\label{ir_excess}
\end{figure*}

To select new YSO candidates in the studied regions we used photometric and variability criteria. First, from the near-infrared (J-H)/(H-K) color - color diagram of each cluster we selected all stars which are at least 3$\sigma$ distant from the reddening line that marks the colors of dwarf stars. The list thereby obtained of 90 stars in all clusters was cross-matched with GLIMPSE measurements. Forty eight of them have photometry from GLIMPSE, and only these objects are proceeded for further inspection. Their coordinates and magnitudes are listed in Table~\ref{YSO_candidates} and Fig.~\ref{ir_excess} shows their [$K_{\rm S} - [3.6]$],[$[3.6] - [4.5]$] colors. The objects with [$K_{\rm S} - [3.6]]> 0.5$ or [$[3.6] - [4.5]]> 0.5$ magnitudes are considered as a most probable class I and class II YSOs . These limits are set in order to avoid selecting objects that are more likely class III objects or normal stars (dashed red line in Fig.~\ref{ir_excess}). 

\begin{deluxetable}{llccccccccc}
\tabletypesize{\tiny} 
\setlength{\tabcolsep}{0.03in} 
\tablewidth{0pt}
\tablecaption{Infrared magnitudes of the YSO candidates.}
\tablehead{
\colhead{Name}            &
\colhead{GLIMPSE Id.}            &
\colhead{RA}  &
\colhead{DEC}    &
\colhead{J}    &
\colhead{H} &
\colhead{Ks}&
\colhead{3.6}&
\colhead{4.5}&
\colhead{5.8}    &
\colhead{8.0} \\
\colhead{(1)}&
\colhead{(2)}&
\colhead{(3)}&
\colhead{(4)}&
\colhead{(5)}&
\colhead{(6)}&
\colhead{(7)}&
\colhead{(8)}&
\colhead{(9)}&
\colhead{(10)}&
\colhead{(11)}\\
}
\startdata
CL010\,ysoc1	&	G298.2663+00.7354	&	182.956177	&	-61.77687	&	19.06$\pm$0.03	&	16.55$\pm$0.01	&	15.02$\pm$0.04	&	13.39$\pm$0.11	&	12.63$\pm$0.13	&	11.84$\pm$0.13	&	10.63$\pm$0.06	\\
CL012\,ysoc1	&	G299.3849-00.2418	&	185.054530	&	-62.89840	&	17.65$\pm$0.03	&	16.43$\pm$0.05	&	14.94$\pm$0.03	&	14.00$\pm$0.09	&	14.21$\pm$0.18	&	\nodata	 	\nodata	&	\nodata	 	\nodata	\\
CL012\,ysoc2	&	G299.3948-00.2348	&	185.078100	&	-62.89268	&	16.55$\pm$0.01	&	15.29$\pm$0.01	&	14.41$\pm$0.02	&	13.37$\pm$0.08	&	12.81$\pm$0.11	&	12.00$\pm$0.20	&	\nodata	 	\nodata	\\
CL012\,ysoc3	&	G299.3814-00.2271	&	185.050720	&	-62.88340	&	14.48$\pm$0.01	&	13.14$\pm$0.01	&	11.97$\pm$0.01	&	10.24$\pm$0.06	&	9.58$\pm$0.05	&	8.90$\pm$0.11	&	7.76$\pm$0.22	\\
CL012\,ysoc4	&	G299.3805-00.2213	&	185.050323	&	-62.87749	&	17.38$\pm$0.02	&	15.84$\pm$0.01	&	14.60$\pm$0.01	&	12.91$\pm$0.05	&	12.40$\pm$0.09	&	11.77$\pm$0.14	&	11.01$\pm$0.26	\\
CL013\,ysoc1	&	G300.3356-00.2167	&	187.138382	&	-62.97299	&	16.47$\pm$0.01	&	12.69$\pm$0.01	&	10.70$\pm$0.01	&	9.23$\pm$0.04	&	9.06$\pm$0.04	&	8.56$\pm$0.05	&	8.55$\pm$0.06	\\
CL013\,ysoc2	&	G300.3475-00.2043	&	187.166850	&	-62.96175	&	20.25$\pm$0.21	&	16.39$\pm$0.01	&	14.33$\pm$0.01	&	12.70$\pm$0.07	&	12.39$\pm$0.10	&	12.19$\pm$0.19	&	\nodata	 	\nodata	\\
CL013\,ysoc3	&	G300.3450-00.2021	&	187.161850	&	-62.95934	&	19.88$\pm$0.16	&	16.49$\pm$0.01	&	14.84$\pm$0.02	&	13.60$\pm$0.07	&	13.34$\pm$0.17	&	\nodata	 	\nodata	&	\nodata	 	\nodata	\\
CL013\,ysoc4	&	G300.3485-00.2288	&	187.164460	&	-62.98619	&	\nodata	 	\nodata	&	17.05$\pm$0.02	&	14.58$\pm$0.01	&	12.73$\pm$0.07	&	12.5$\pm$0.12	&	12.42$\pm$0.22	&	\nodata	 	\nodata	\\
CL013\,ysoc5	&	G322.000+00.6165	&	187.143190	&	-62.96543	&	\nodata	 	\nodata	&	18.16$\pm$0.04	&	16.02$\pm$0.04	&	14.28$\pm$0.12	&	14.1$\pm$0.21	&	\nodata	 	\nodata	&	\nodata	 	\nodata	\\
CL013\,ysoc6	&	G300.3240-00.1985	&	187.116580	&	-62.95386	&	17.41$\pm$0.02	&	14.16$\pm$0.01	&	11.89$\pm$0.01	&	9.29$\pm$0.06	&	7.90$\pm0.07$	&	6.96$\pm$0.03	&	5.95$\pm$0.02	\\
CL013\,ysoc7	&	G300.3290-00.2046	&	187.126240	&	-62.96040	&	\nodata	 	\nodata	&	19.70$\pm$0.13	&	15.57$\pm$0.03	&	12.39$\pm$0.07	&	11.70$\pm$0.08	&	11.28$\pm$0.08	&	11.47$\pm$0.14	\\
CL013\,ysoc8	&	G300.3420-00.2299	&	187.149780	&	-62.98669	&	\nodata	 	\nodata	&	18.14$\pm$0.05	&	15.48$\pm$0.03	&	13.64$\pm$0.09	&	13.30$\pm$0.13	&	\nodata	 	\nodata	&	\nodata	 	\nodata	\\
CL013\,ysoc9	&	G300.3299-00.2036	&	187.128720	&	-62.95964	&	\nodata	 	\nodata	&	17.77$\pm$0.03	&	15.93$\pm$0.04	&	13.59$\pm$0.09	&	13.23$\pm$0.17	&	\nodata	 	\nodata	&	\nodata	 	\nodata	\\
CL059\,ysoc1	&	G331.2413+01.0751	&	241.455527	&	-50.79151	&	16.22$\pm$0.20	&	13.71$\pm$0.10	&	12.64$\pm$0.05	&	11.86$\pm$0.06	&	11.86$\pm$0.10	&	11.11$\pm$0.15	&	\nodata	 	\nodata	\\
CL059\,ysoc2	&	G331.2560+01.0600	&	241.489030	&	-50.79288	&	\nodata	 	\nodata	&	18.11$\pm$0.06	&	15.72$\pm$0.03	&	13.69$\pm$0.12	&	13.23$\pm$0.18	&	\nodata	 	\nodata	&	\nodata	 	\nodata	\\
CL059\,ysoc3	&	G331.2624+01.0576	&	241.498900	&	-50.79050	&	\nodata	 	\nodata	&	15.27$\pm$0.04	&	12.80$\pm$0.02	&	10.84$\pm$0.05	&	10.55$\pm$0.06	&	10.19$\pm$0.09	&	\nodata	 	\nodata	\\
CL059\,ysoc4	&	G331.2632+01.0488	&	241.509160	&	-50.79656	&	\nodata	 	\nodata	&	17.32$\pm$0.05	&	14.54$\pm$0.04	&	10.37$\pm$0.07	&	9.40$\pm$0.12	&	7.42$\pm$0.04	&	5.95$\pm$0.04	\\
CL059\,ysoc5	&	G331.2534+01.0791	&	241.465600	&	-50.78044	&	18.44$\pm$0.08	&	15.02$\pm$0.01	&	13.29$\pm$0.02	&	12.13$\pm$0.14	&	12.10$\pm$0.17	&	\nodata	 	\nodata	&	\nodata	 	\nodata	\\
CL059\,ysoc6	&	G331.2508+01.0751	&	241.466740	&	-50.78527	&	\nodata	 	\nodata	&	15.91$\pm$0.02	&	12.10$\pm$0.02	&	9.50$\pm$0.04	&	9.03$\pm$0.06	&	8.39$\pm$0.07	&	8.48$\pm$0.12	\\
CL059\,ysoc7	&	G331.2515+01.0714	&	241.471480	&	-50.78754	&	\nodata	 	\nodata	&	17.20$\pm$0.04	&	13.63$\pm$0.02	&	10.77$\pm$0.15	&	10.14$\pm$0.08	&	9.69$\pm$0.17	&	\nodata	 	\nodata	\\
CL059\,ysoc8	&	G331.2391+01.0533	&	241.476060	&	-50.80931	&	14.70$\pm$0.01	&	12.62$\pm$0.01	&	11.25$\pm$0.02	&	9.14$\pm$0.06	&	8.43$\pm$0.07	&	7.43$\pm$0.04	&	6.34$\pm$0.10	\\
DBS93\,ysoc1	&	G322.1434+00.6397	&	229.617635	&	-56.64518	&	17.32$\pm$0.10	&	12.20$\pm$0.03	&	9.76$\pm$0.03	&	8.33$\pm$0.03	&	8.20$\pm$0.03	&	7.70$\pm$0.02	&	7.78$\pm$0.06	\\
DBS93\,ysoc2	&	G322.1399+00.6073	&	229.643649	&	-56.67440	&	16.83$\pm$0.10	&	14.31$\pm$0.06	&	12.52$\pm$0.04	&	11.35$\pm$0.04	&	11.09$\pm$0.06	&	10.88$\pm$0.10	&	\nodata	 	\nodata	\\
DBS93\,ysoc3	&	G322.1624+00.6065	&	229.679138	&	-56.66308	&	14.45$\pm$0.04	&	11.55$\pm$0.03	&	10.20$\pm$0.05	&	9.14$\pm$0.03	&	8.99$\pm$0.03	&	8.59$\pm$0.03	&	8.43$\pm$0.05	\\
DBS93\,ysoc4	&	G322.1493+00.6468	&	229.619956	&	-56.63598	&	17.98$\pm$0.10	&	14.93$\pm$0.09	&	10.59$\pm$0.02	&	8.10$\pm$0.03	&	7.49$\pm$0.03	&	6.87$\pm$0.03	&	7.06$\pm$0.10	\\
DBS93\,ysoc5	&	G322.1606+00.6315	&	229.652039	&	-56.64293	&	18.02$\pm$0.08	&	15.33$\pm$0.04	&	12.52$\pm$0.01	&	10.51$\pm$0.10	&	9.94$\pm$0.12	&	\nodata	 	\nodata	&	\nodata	 	\nodata	\\
DBS93\,ysoc6	&	G322.1378+00.6453	&	229.603586	&	-56.64346	&	17.78$\pm$0.10	&	14.46$\pm$0.06	&	12.34$\pm$0.03	&	10.94$\pm$0.04	&	10.65$\pm$0.06	&	10.21$\pm$0.06	&	9.70$\pm$0.23	\\
DBS93\,ysoc7	&	G322.1834+00.6391	&	229.679860	&	-56.62423	&	15.40$\pm$0.08	&	11.68$\pm$0.03	&	9.98$\pm$0.03	&	8.82$\pm$0.03	&	8.69$\pm$0.05	&	8.33$\pm$0.05	&	\nodata	 	\nodata	\\
DBS93\,ysoc8	&	G322.1870+00.6286	&	229.695662	&	-56.63129	&	16.31$\pm$0.10	&	12.98$\pm$0.04	&	11.25$\pm$0.03	&	9.96$\pm$0.04	&	9.84$\pm$0.05	&	9.31$\pm$0.05	&	8.93$\pm$0.09	\\
DBS93\,ysoc9	&	G322.1374+00.6164	&	229.631014	&	-56.66809	&	17.71$\pm$0.10	&	15.10$\pm$0.11	&	12.77$\pm$0.03	&	11.35$\pm$0.04	&	11.14$\pm$0.06	&	11.00$\pm$0.15	&	\nodata	 	\nodata	\\
DBS93\,ysoc10	&	G322.1666+00.6048	&	229.687341	&	-56.66238	&	17.37$\pm$0.10	&	13.96$\pm$0.05	&	12.53$\pm$0.03	&	11.51$\pm$0.03	&	11.47$\pm$0.06	&	10.89$\pm$0.09	&	\nodata	 	\nodata	\\
DBS93\,ysoc11	&	G322.1539+00.6228	&	229.650269	&	-56.65396	&	18.14$\pm$0.05	&	15.27$\pm$0.02	&	13.36$\pm$0.02	&	11.29$\pm$0.10	&	10.55$\pm$0.16	&	\nodata	 	\nodata	&	\nodata	 	\nodata	\\
DBS100\,ysoc1	&	G332.8445-00.5846	&	245.106613	&	-50.89970	&	18.12$\pm$0.04	&	14.25$\pm$0.01	&	12.35$\pm$0.01	&	10.87$\pm$0.07	&	10.87$\pm$0.09	&	10.67$\pm$0.21	&	\nodata	 	\nodata	\\
DBS100\,ysoc2	&	G332.8571-00.5875	&	245.123965	&	-50.89287	&	16.96$\pm$0.10	&	15.92$\pm$0.10	&	13.28$\pm$0.04	&	8.76$\pm$0.04	&	7.49$\pm$0.05	&	6.17$\pm$0.03	&	5.27$\pm$0.03	\\
DBS100\,ysoc3	&	G332.8418-00.5841	&	245.102951	&	-50.90125	&	16.17$\pm$0.01	&	12.72$\pm$0.01	&	10.96$\pm$0.02	&	9.65$\pm$0.04	&	9.63$\pm$0.06	&	9.27$\pm$0.06	&	9.60$\pm$0.11	\\
DBS100\,ysoc4	&	G332.8689-00.5850	&	245.134457	&	-50.88289	&	15.81$\pm$0.10	&	13.58$\pm$0.08	&	11.23$\pm$0.04	&	8.95$\pm$0.14	&	8.02$\pm$0.13	&	7.08$\pm$0.06	&	5.71$\pm$0.04	\\
DBS100\,ysoc5	&	G332.8482-00.5897	&	245.116364	&	-50.90090	&	18.19$\pm$0.07	&	15.36$\pm$0.02	&	13.91$\pm$0.02	&	12.39$\pm$0.10	&	12.16$\pm$0.18	&	\nodata	 	\nodata	&	\nodata	 	\nodata	\\
DBS100\,ysoc6	&	G332.8383-00.5887	&	245.104156	&	-50.90699	&	15.90$\pm$0.01	&	13.19$\pm$0.01	&	11.90$\pm$0.01	&	10.77$\pm$0.09	&	10.71$\pm$0.07	&	10.46$\pm$0.08	&	\nodata	 	\nodata	\\
DBS100\,ysoc7	&	G332.8399-00.5876	&	245.105026	&	-50.90490	&	19.23$\pm$0.10	&	16.55$\pm$0.02	&	15.31$\pm$0.03	&	13.59$\pm$0.13	&	13.05$\pm$0.15	&	\nodata	 	\nodata	&	\nodata	 	\nodata	\\
DBS100\,ysoc8	&	G332.8347-00.5848	&	245.095718	&	-50.90691	&	17.95$\pm$0.14	&	15.34$\pm$0.02	&	14.14$\pm$0.01	&	12.58$\pm$0.10	&	12.26$\pm$0.12	&	\nodata	 	\nodata	&	\nodata	 	\nodata	\\
DBS100\,ysoc9	&	G332.8395-00.5964	&	245.114143	&	-50.91167	&	11.66$\pm$0.04	&	9.12$\pm$0.04	&	7.94$\pm$0.02	&	7.03$\pm$0.04	&	6.94$\pm$0.05	&	6.58$\pm$0.03	&	6.57$\pm$0.03	\\
DBS100\,ysoc10	&	G332.8716-00.5631	&	245.113112	&	-50.86538	&	15.71$\pm$0.09	&	13.47$\pm$0.05	&	12.02$\pm$0.04	&	10.37$\pm$0.05	&	9.79$\pm$0.05	&	9.12$\pm$0.04	&	7.63$\pm$0.06	\\
DBS130\,ysoc1	&	G305.2451+00.0056	&	197.921080	&	-62.77509	&	14.43$\pm$0.05	&	10.82$\pm$0.03	&	9.07$\pm$0.03	&	7.73$\pm$0.04	&	7.61$\pm$0.04	&	7.16$\pm$0.02	&	7.05$\pm$0.06	\\
DBS130\,ysoc2	&	G305.2646-00.0050	&	197.965347	&	-62.78419	&	18.13$\pm$0.02	&	14.77$\pm$0.01	&	13.13$\pm$0.01	&	12.08$\pm$0.11	&	11.99$\pm$0.27	&	\nodata	 	\nodata	&	\nodata	 	\nodata	\\
DBS130\,ysoc3	&	G305.2530+00.0034	&	197.938602	&	-62.77666	&	17.32$\pm$0.05	&	14.89$\pm$0.07	&	12.79$\pm$0.02	&	11.32$\pm$0.09	&	11.08$\pm$0.10	&	\nodata	 	\nodata	&	\nodata	 	\nodata	\\
DBS130\,ysoc4	&	G305.2845-00.0122	&	198.010098	&	-62.78972	&	15.33$\pm$0.05	&	13.63$\pm$0.07	&	11.65$\pm$0.03	&	10.30$\pm$0.07	&	10.09$\pm$0.08	&	9.59$\pm$0.07	&	\nodata	 	\nodata	\\
DBS130\,ysoc5	&	G305.2641+00.0009	&	197.963330	&	-62.77818	&	18.33$\pm$0.04	&	16.33$\pm$0.01	&	14.73$\pm$0.03	&	12.69$\pm$0.12	&	11.85$\pm$0.26	&	\nodata	 	\nodata	&	\nodata	 	\nodata	\\
\enddata
\label{YSO_candidates}
\end{deluxetable}

\begin{figure}
\epsscale{1.0}
\plotone{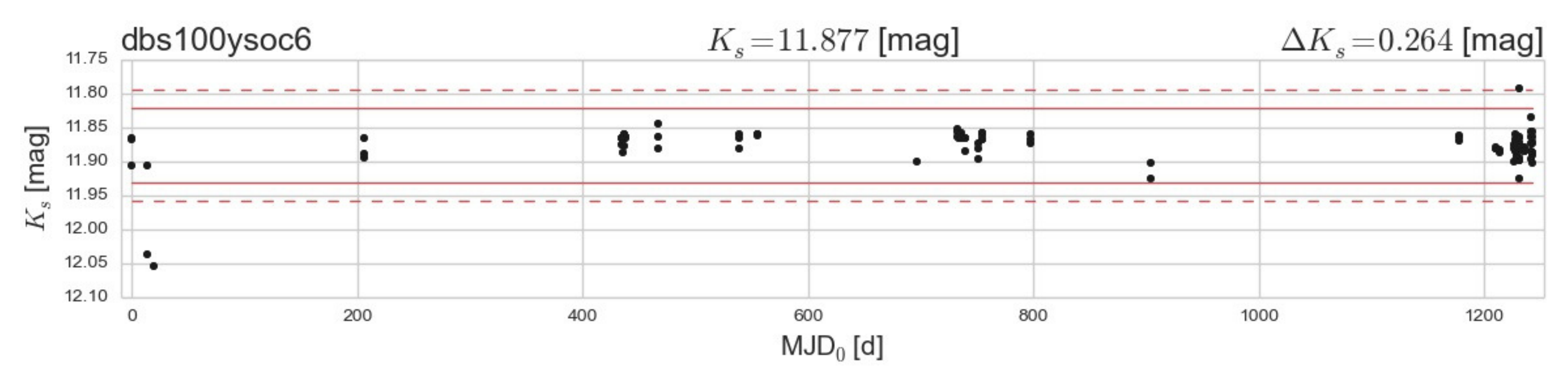}
\plotone{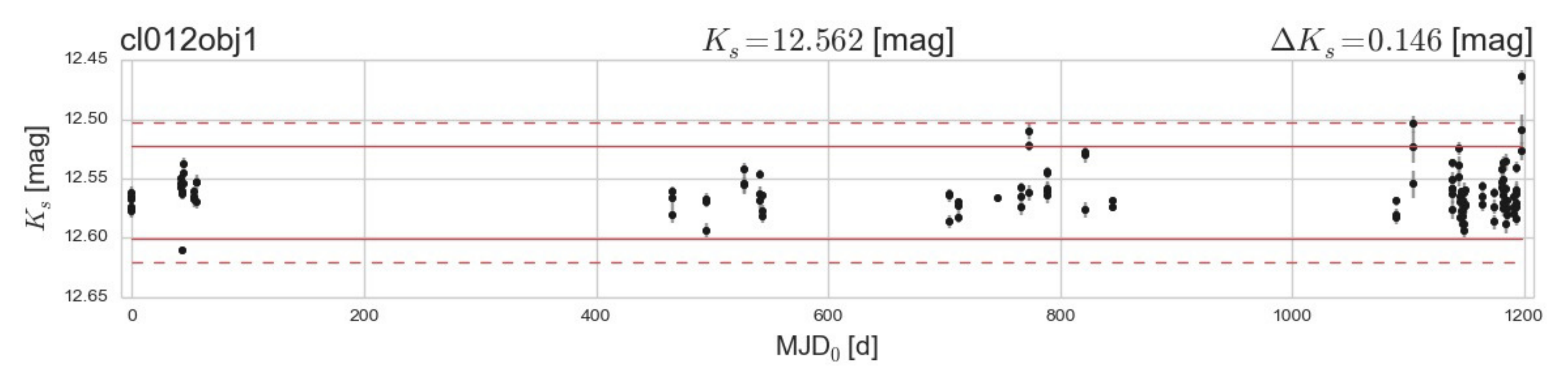}
\plotone{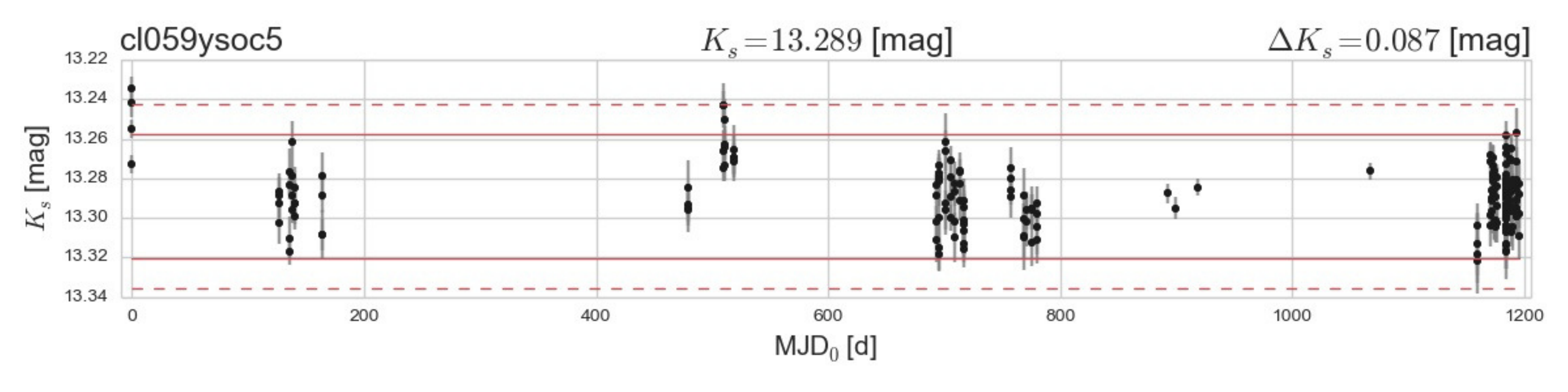}
\plotone{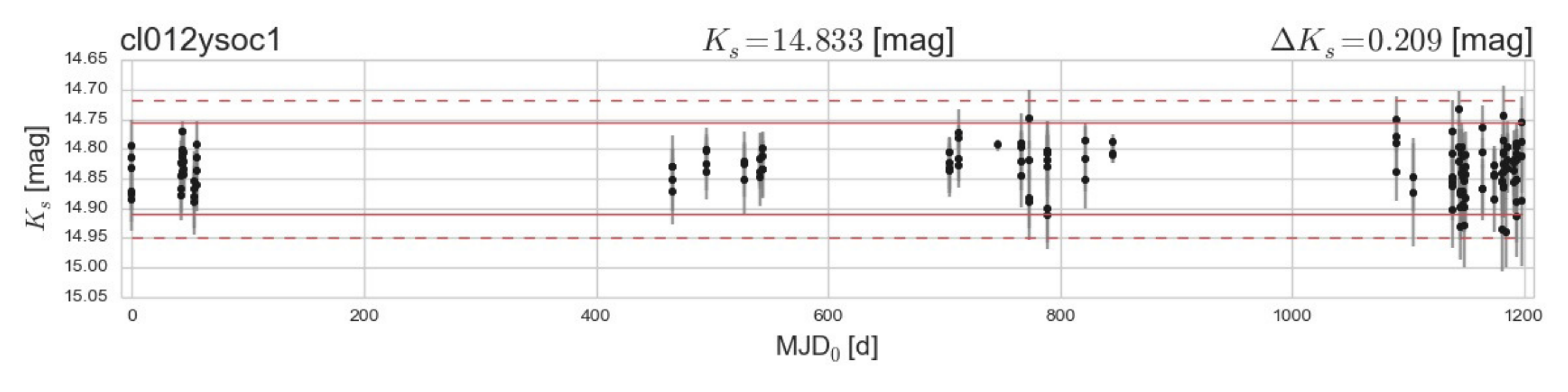}
\vspace{0.1cm}
\caption{Examples of $K_{\rm S}$ magnitude vs MJD of non variable YSO candidates. The solid and dashed red lines mark the 2 and 3 $\sigma $ dispersion the light curve, respectively.} 
\label{nonvar}
\end{figure}

\begin{figure}
\epsscale{0.7}
\plotone{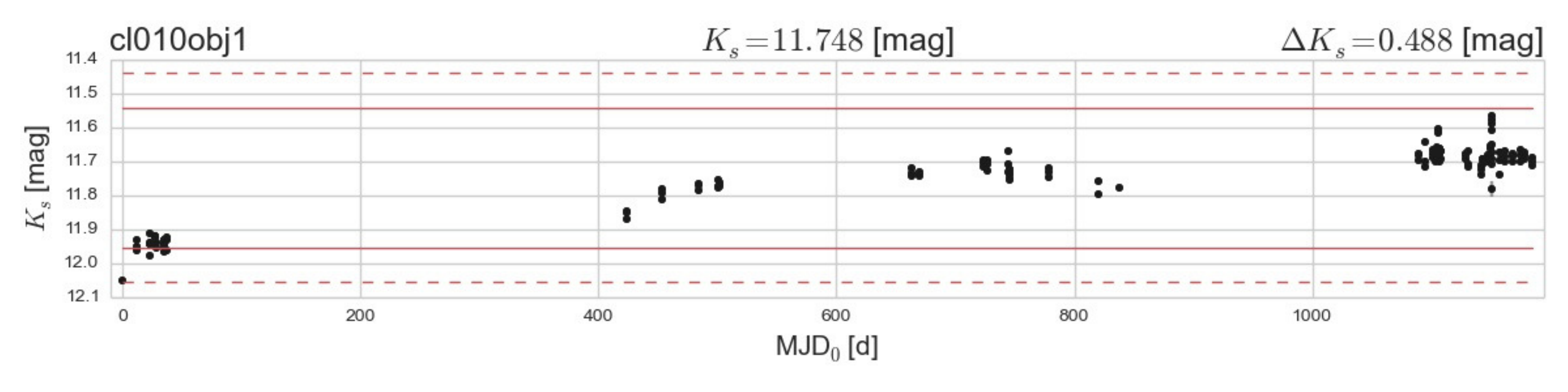}
\plotone{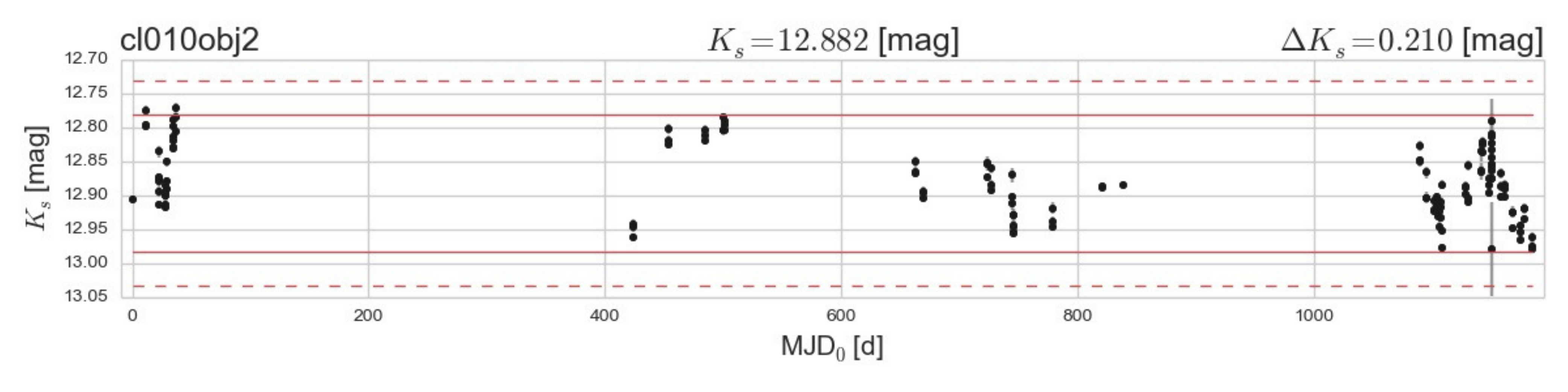}
\plotone{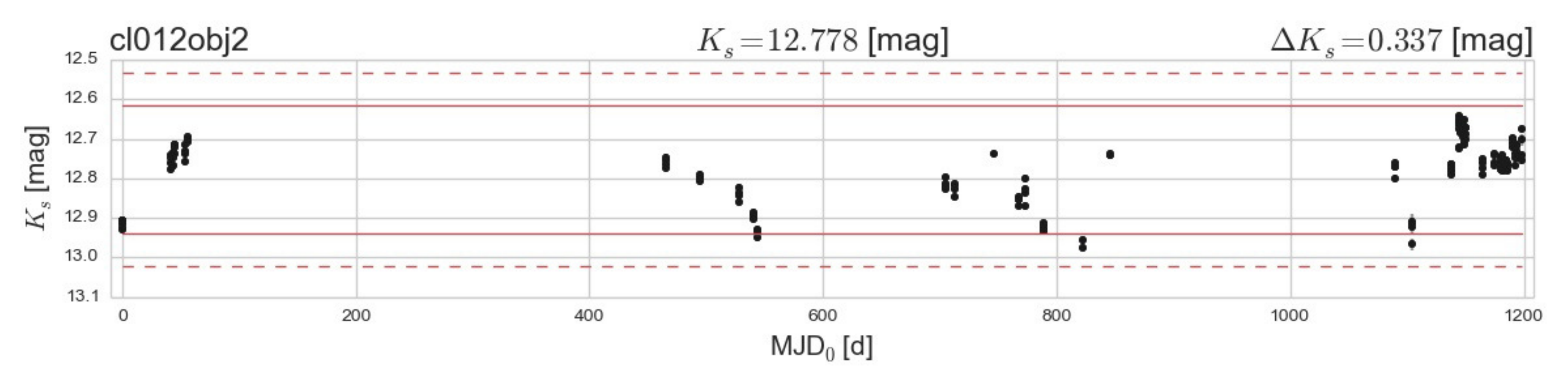}
\plotone{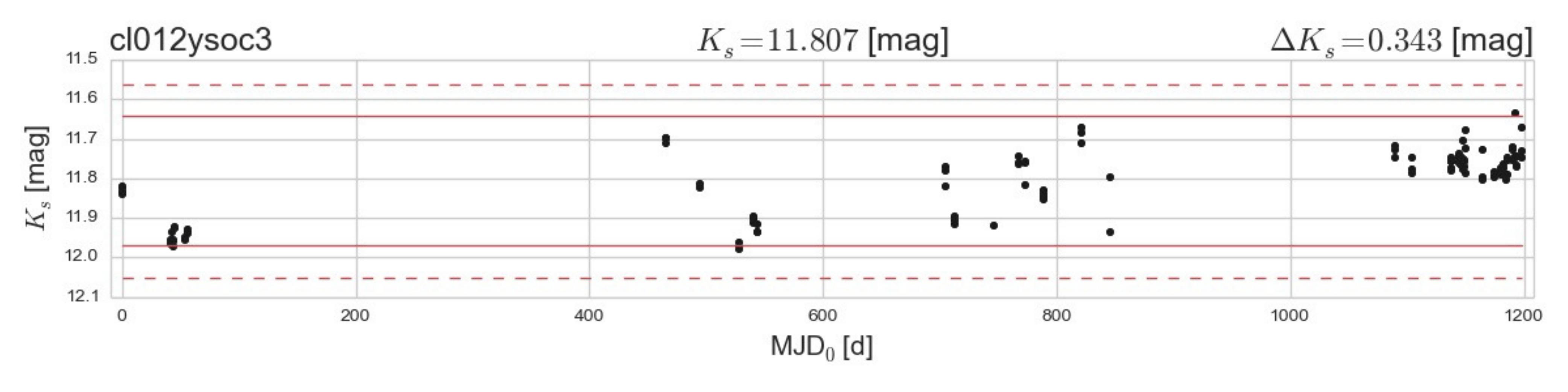}
\plotone{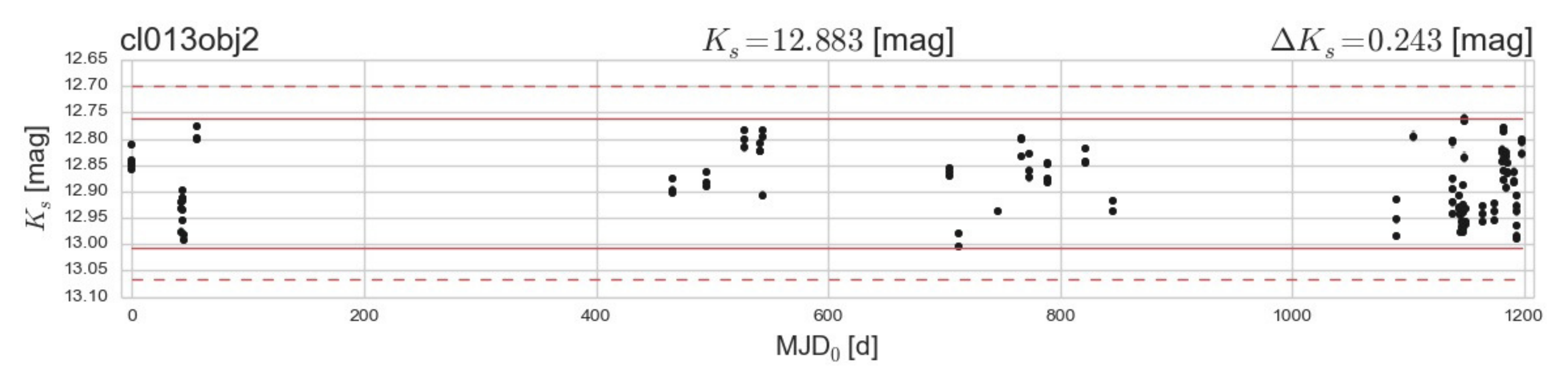}
\plotone{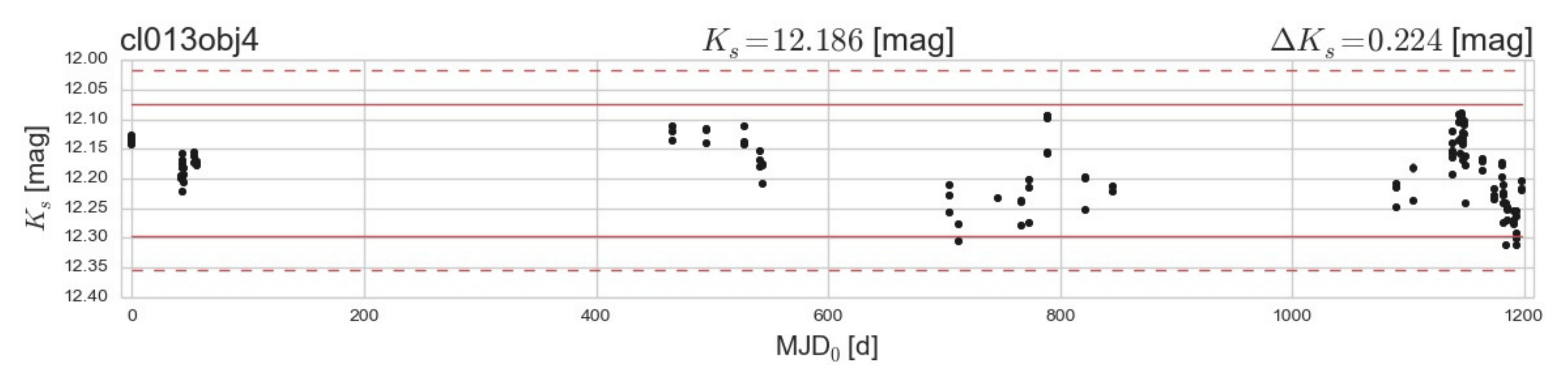}
\plotone{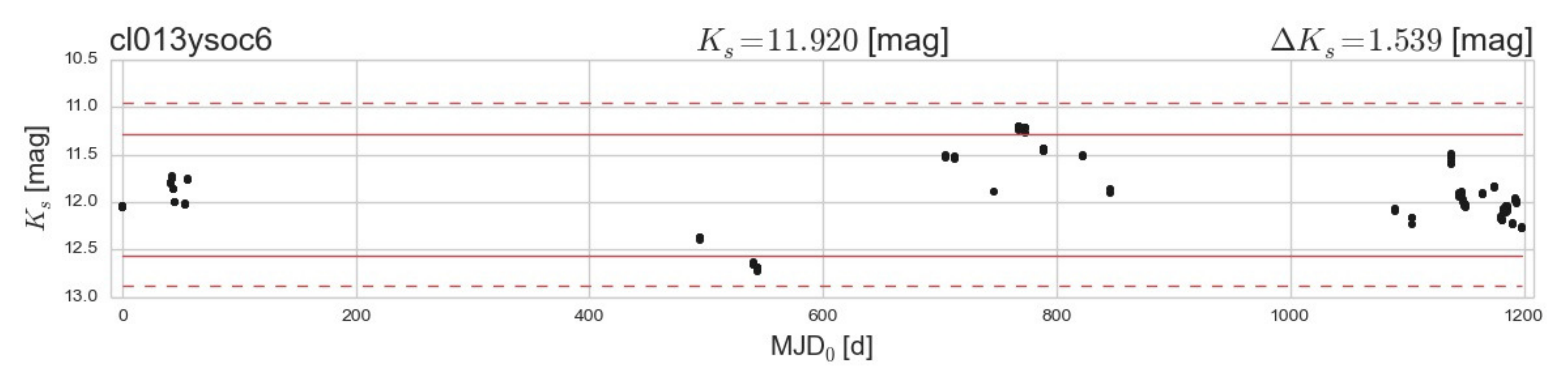}
\plotone{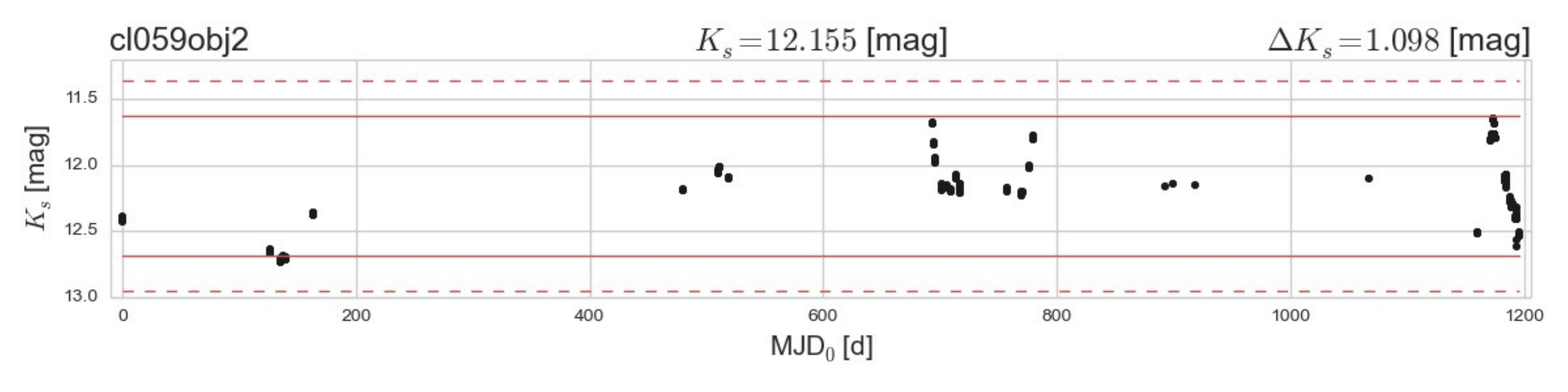}
\vspace{0.1cm}
\caption{Examples of $K_{\rm S}$ magnitude vs MJD of variable YSO candidates. The symbols are the same as in Fig.~\ref{nonvar}.} 
\label{var1}
\end{figure}

\begin{figure}
\epsscale{0.7}
\plotone{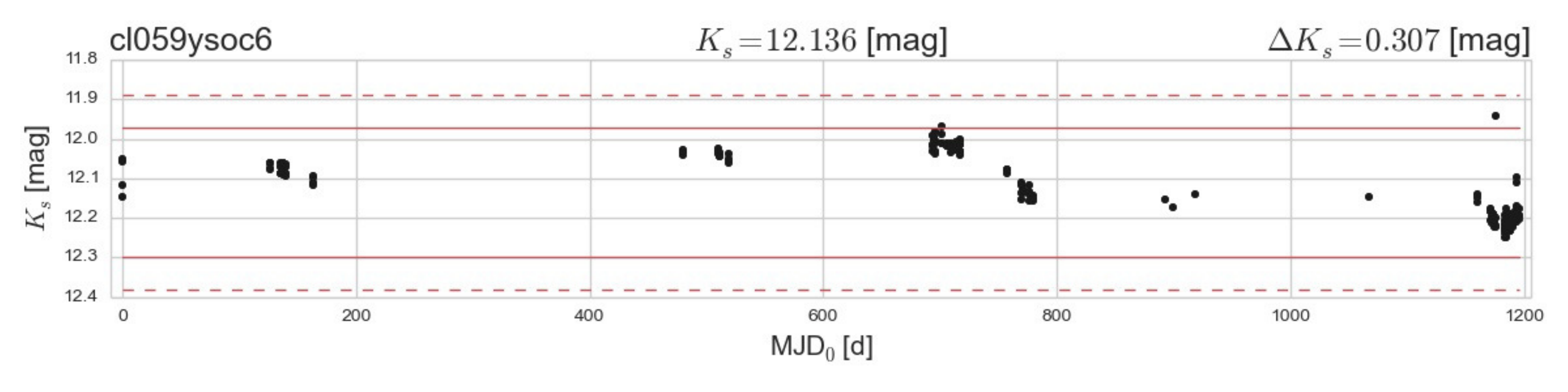}
\plotone{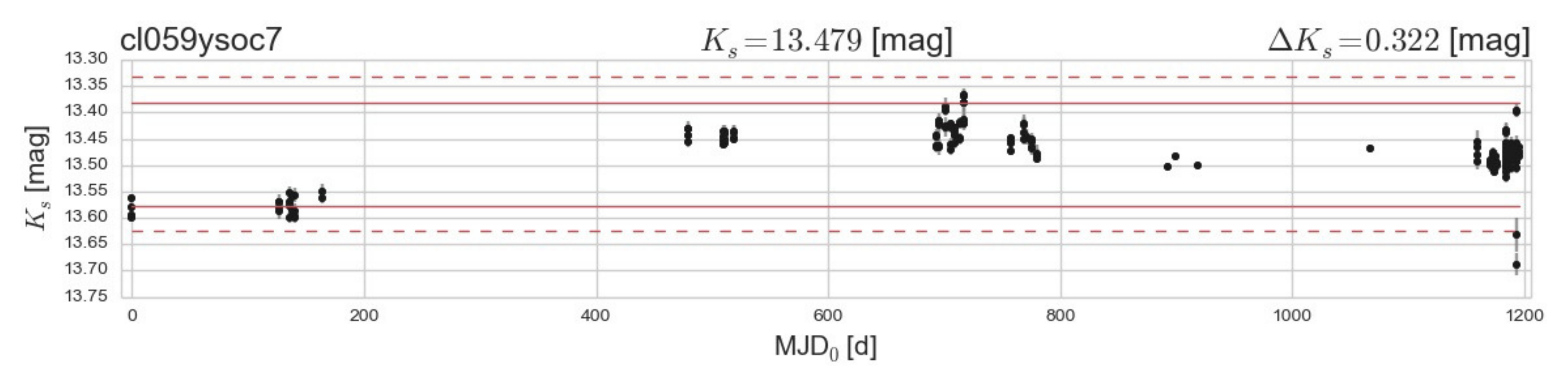}
\plotone{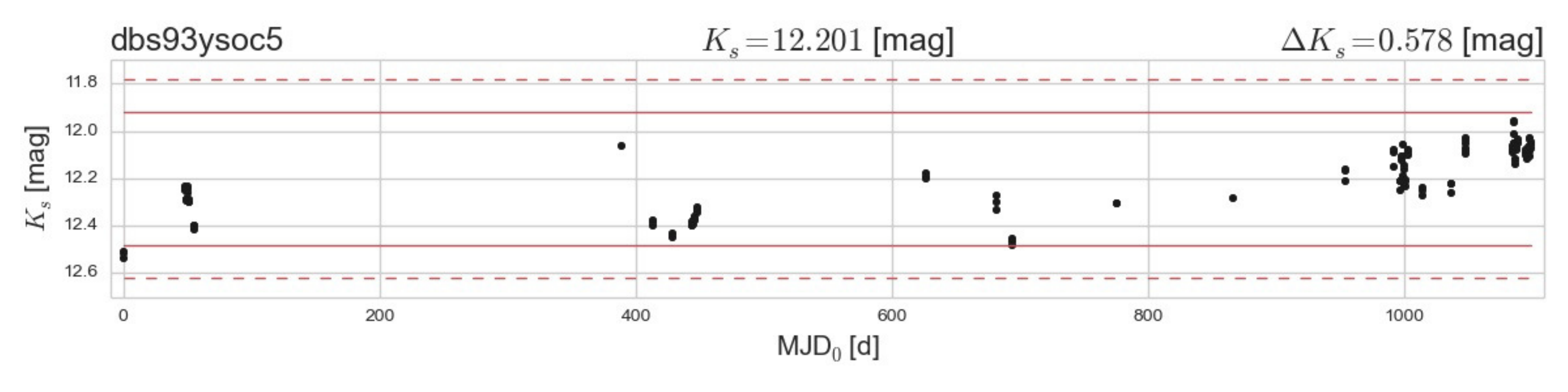}
\plotone{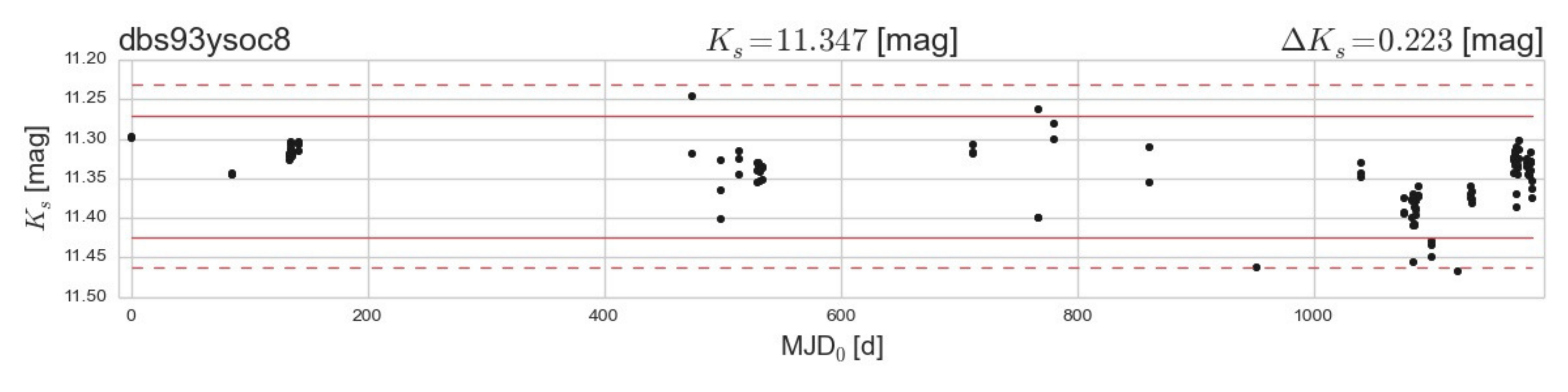}
\plotone{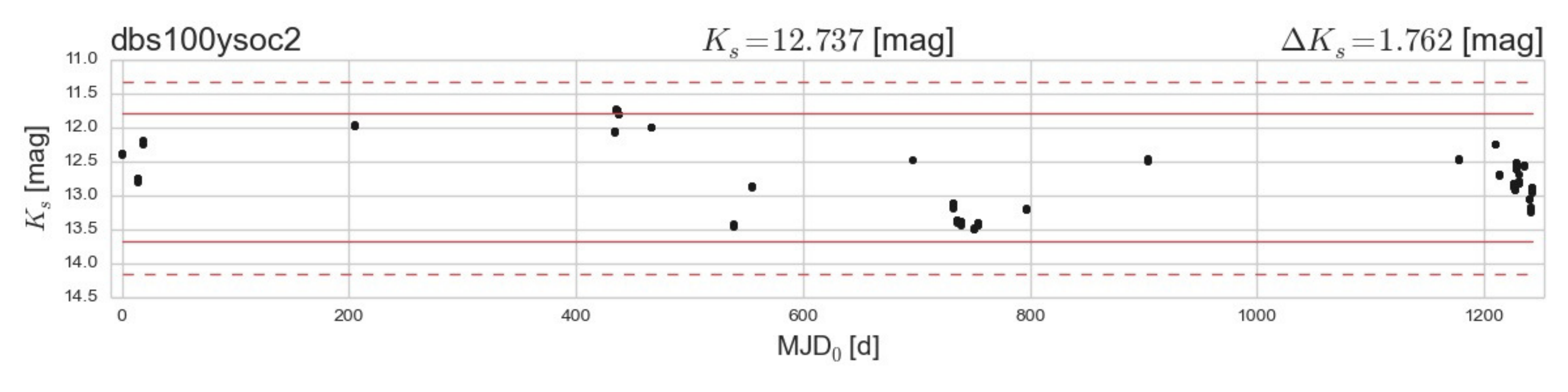}
\plotone{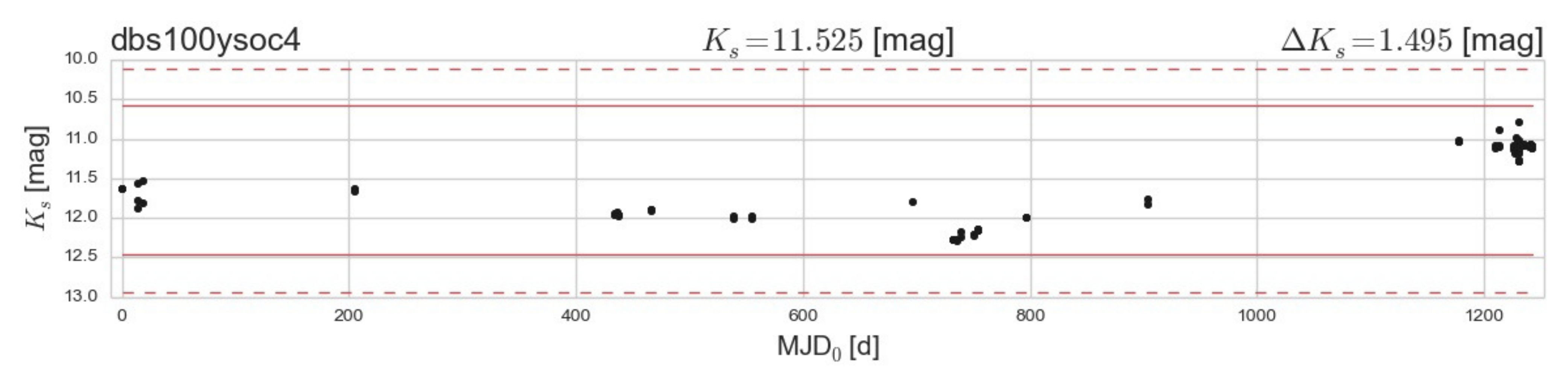}
\plotone{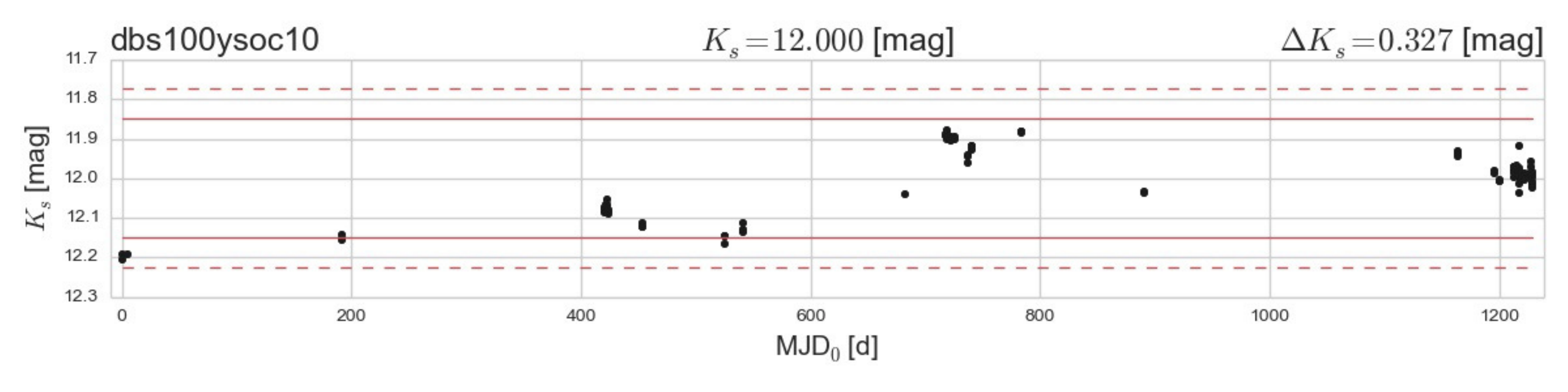}
\plotone{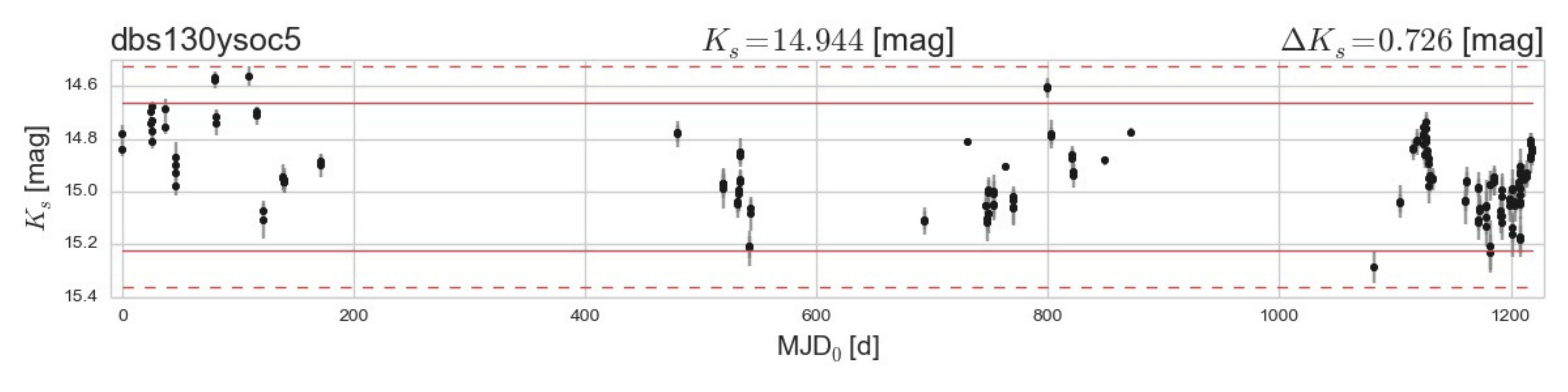}
\vspace{0.1cm}
\caption{Examples of $K_{\rm S}$ magnitude vs MJD of variable YSO candidates. The symbols are the same as in Fig.~\ref{nonvar}.} 
\label{var2}
\end{figure}

{

All available $K_{\rm S}$ magnitude in VISTA Science Archive (VSA) Data Release 4 (DR4, four year database, up to 30 September 2013, http://horus.roe.ac.uk/vsa/index.html) with grades A and B (e.g. observed within optimal sky conditions) are retrieved to check the variability of the above selected YSO candidates, together with spectroscopically confirmed candidates. The level of variability seen in normal, non-variable stars is estimated to be below 0.1 mag at 12$<$Ks$<$16 using apermag3 (2\arcsecspace diameter aperture) in the tile catalogs, but we put a 0.2 mag conservative limit marking the errors of the photometry and transformation to the standard system. The saturated stars, the objects with close companions (blending) and those with large photometric errors, 10 in total are removed.  To analyze the rest of them we compute a set of variability indexes, namely Stetson J and K indexes (Stetson 1996), the $\eta$ index (von Neumann 1941), the chi square test $\chi^{2}$ (Rebull et all. 2015), the Small Kurtosis $\kappa$ (Richards et al. 2011) and $\frac{\mu}{\sigma}$ (e.g. the ratio between the average $K_{\rm S}$ magnitude from light curve over the standard deviation of the data). Then we used an Unsupervised Clustering Algorithm, which identify patterns among the values of these indexes, separating populations of objects with similar features. Thus, two groups of objects are defined, one with significant amplitudes (in general, greater than 0.2 mag in $K_{\rm S}$), which shows long and short term variability in the time domain, and another group of sources that the variability is not very significant, can be confused with noise and thus is uncertain.   

Fig.~\ref{nonvar},  Fig.~\ref{var1} and Fig.~\ref{var2} show examples of MJD vs. $K_{\rm S}$ magnitudes for the non variable and variable stars, respectively. According to our analysis 57 \%  of the YSO candidates show signatures of infrared variability. Most of the variable stars in our sample show amplitude variations between 0.2 and 0.5 mag, and only six stars have higher amplitudes. Actually, CL059\,Obj2 was taken from the Contreras Pena (2016) list of high amplitude variables, and according to SIMBAD the object CL013\,ysoc6 (2MASS J12282798-6257139) is a YSO candidate; DBS100\,ysoc2 (2MASS J16202975-5053343) is an AGB candidate; DBS100\,ysoc4 (2MASS J16203226-5052584) is a YSO candidate in the list of intrinsically red stars in the Robitaille et al.(2008) paper. We try to determine some periods using different statistic methods, unfortunately, this was not possible on the base of the existing epochs. Nevertheless, according to the light curves and the position in the color-magnitude diagrams, we consider that the stars CL013\,ysoc6; DBS100\,ysoc2; DBS100\,ysoc4 and DBS130\,ysoc5 are most probably semi-regular asymptotic variable stars. It is well known (Robitaille et al. 2008) that the color-cut photometric selections alone can not distinguish between YSO and AGB stars. However, as we show above, the combination with infrared variability analysis can help to solve this problem.

}

\begin{figure}
\epsscale{0.5}
\plottwo{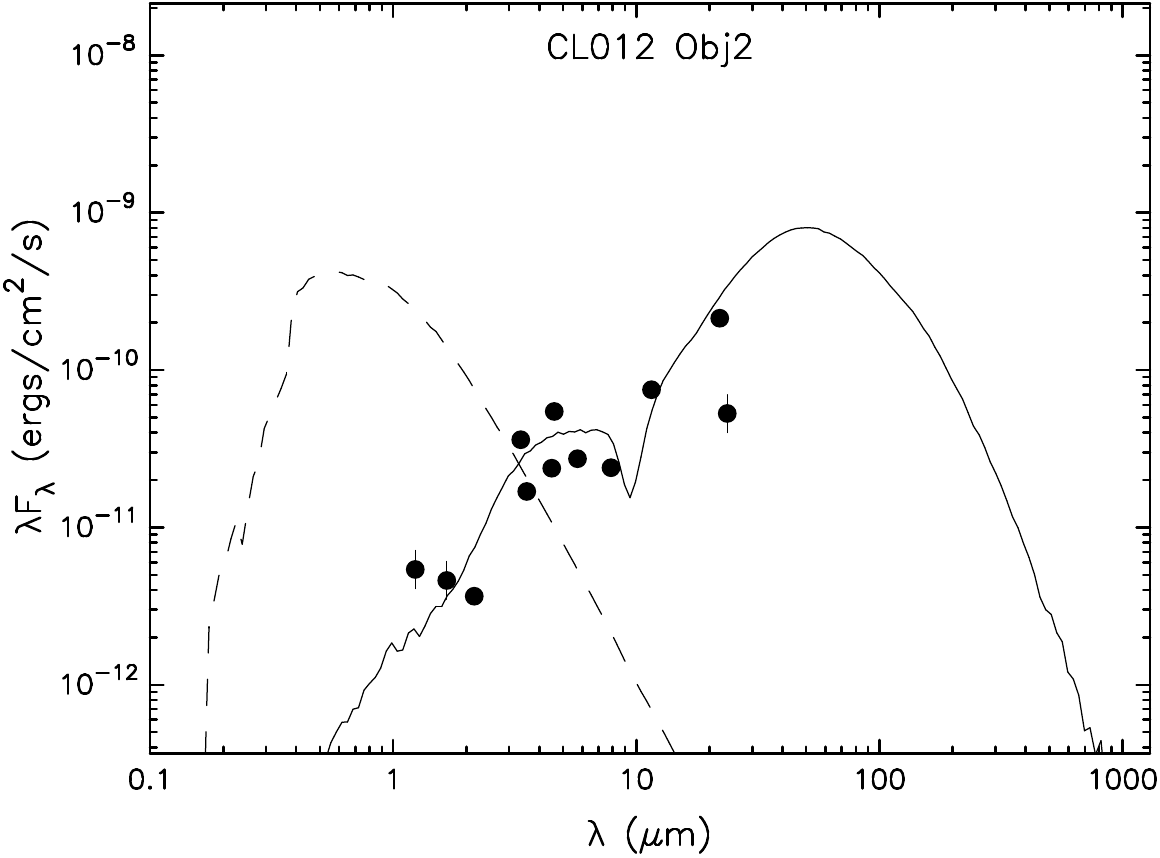}{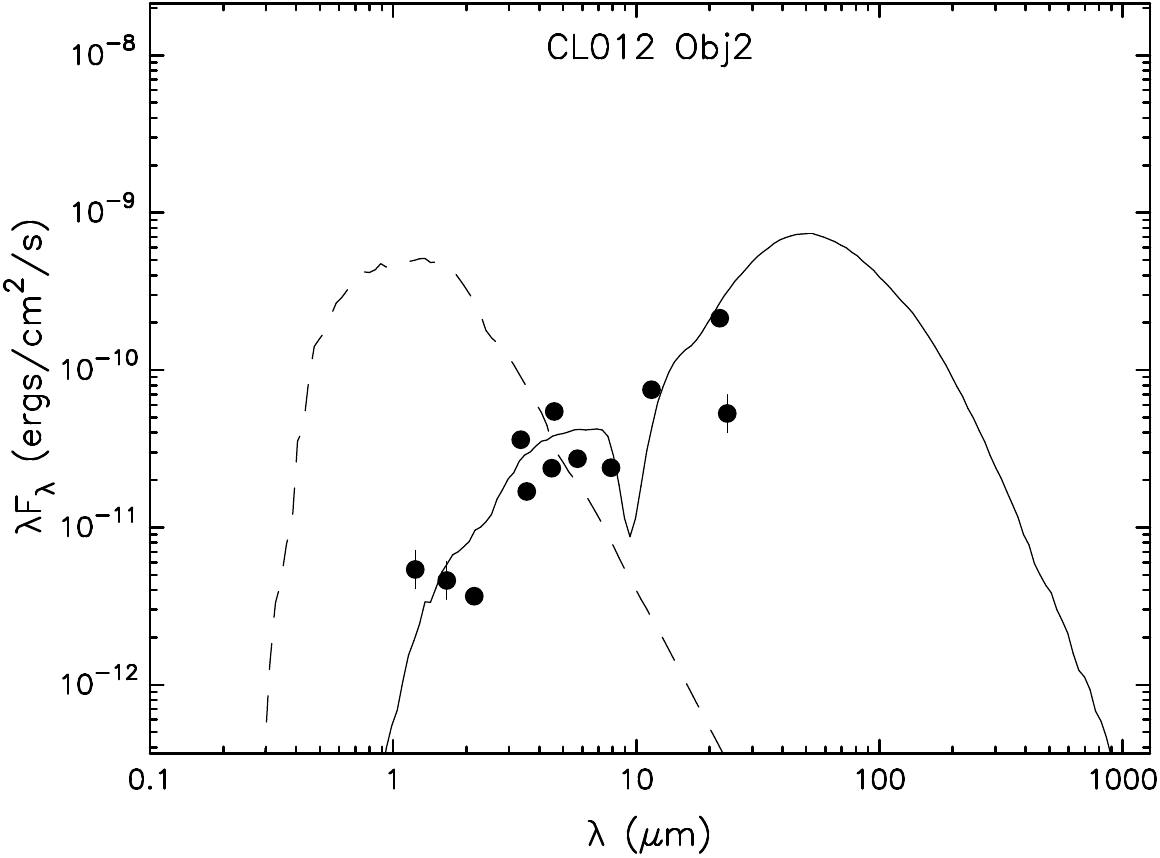}
\plottwo{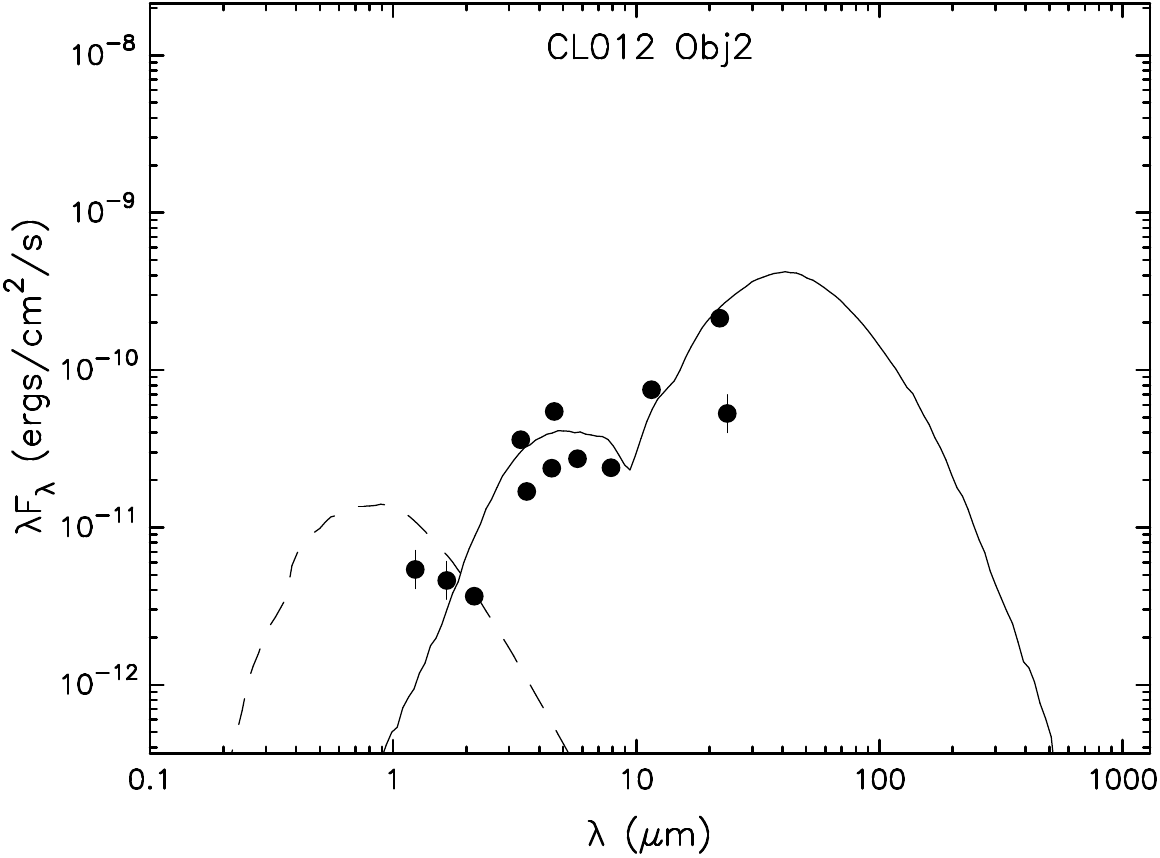}{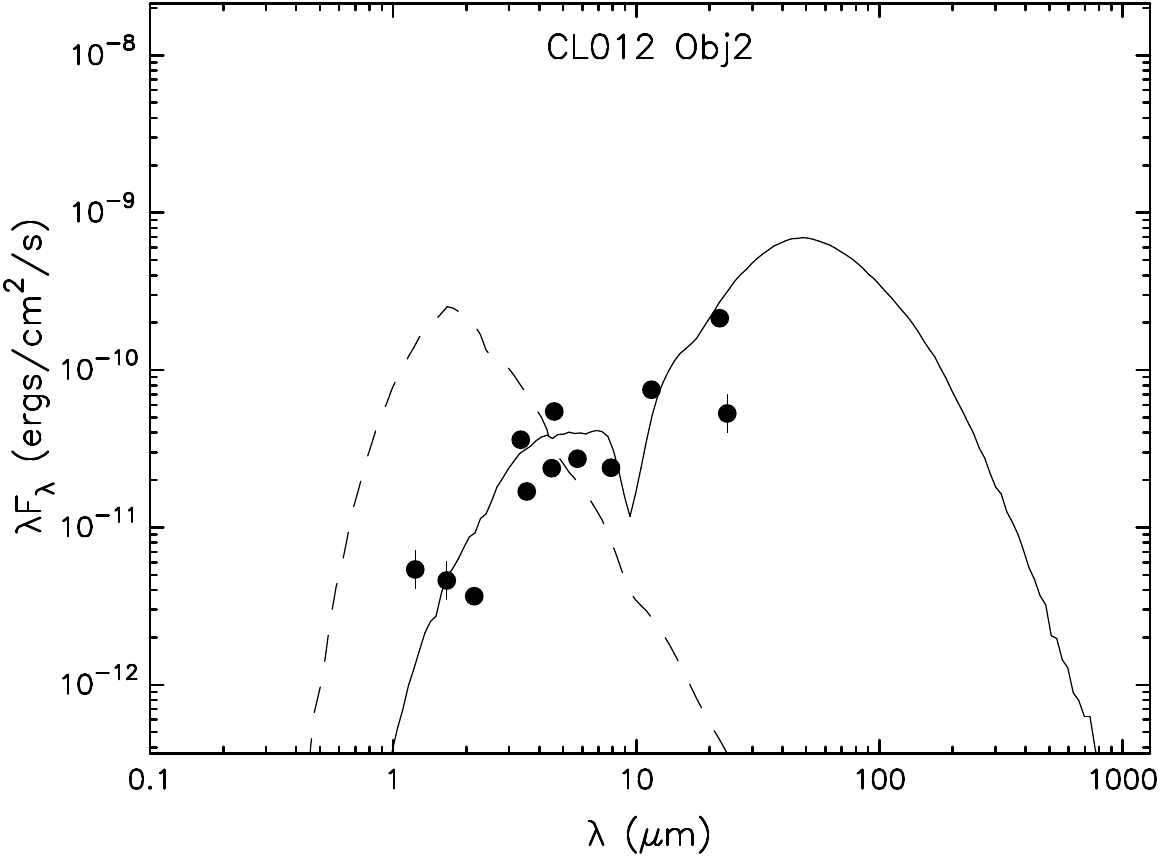}
\caption{Four different SEDs models (solid lines) for the CL\,012 Obj2. The dashed line plots the best fit photometric model.} 
\label{sed}
\end{figure}

{
To model the Spectral Energy Distributions (SED) of the YSO candidates we use the spectral energy distribution models of young stellar objects developed by Robitaille et al. (2008). We have collected existing measurements of these objects, from the VVV, 2MASS, GLIMPSE, WISE and HIGAL catalogs. The respective reddening and distances of each object are determined by photometry and spectroscopy in the Sect.~3, however due to their large  uncertainties, we considered all models that lie within a 5 $\sigma$ of the respective errors. The Robitaille models never give a single result, rather, they give a range of models and a chi-squared parameter, an example is shown in Fig.~\ref{sed} for CL\,012 Obj2. The range of models that give an acceptable chi-squared is usually up to 5 models starting from the best fit model in our case, since the HIGAL magnitudes constrains the longer wavelength range. Thus, we calculate the mean and standard deviation of stellar ages, masses, temperature and luminosity of YSOs from the adopted models, which are tabulated in Table~\ref{sed_table}. Only seven stars are fitted, the rest of the objects do not have enough measurements to construct reliable SEDs. As can be seen from the Table~\ref{sed_table}, six of our sources can be classified as massive young stellar objects, with masses grater that 8$\Msun$, only DBS100\,Obj1 is an intermediate-mass objects. All of the objects are very young. The  stellar temperatures of the sources are ranging from 4400 to 37000 K. Every SED model shows a presence of envelope. Only for DBS100\,Obj1 the presence of a disk is not detected. It's interesting that Cl010\,Obj1 is very massive and has a relatively large amplitude of variability (0.49 mag in $K_S$). Such high variability has rarely been seen in massive YSOs.  

}
	
\begin{deluxetable}{lllllcl}
\tabletypesize{\scriptsize} 
\tablewidth{0pt}
\tablecaption{The stellar ages, masses, temperature and luminosity of YSOs. Note that the errors are also in logarithmic scale.}
\tablehead{
\colhead{Object}            &
\colhead{Log(Age)}  &
\colhead{Log(Mass) (\Msun)}    &
\colhead{Log(T) $(K)$}    &
\colhead{Log(Disk Mass) ({\Msun})} &
\colhead{{Log(L$_{tot}$) (\Lsun})} &
\colhead{Log(Env. Mass) ({\Msun})}   }           
\startdata
CL010\,Obj1		&	4.38$\pm$0.09	& 1.23$\pm$0.05&4.32$\pm$0.14&-0.65$\pm$0.40&4.59$\pm$0.19&3.43$\pm$0.10\\
CL012\,Obj2		&	4.14$\pm$0.52 &	0.82$\pm$0.31&3.69$\pm$0.10&-1.15$\pm$0.54&2.94$\pm$0.82&	1.83$\pm$0.61\\
CL013\,Obj1		&	5.85$\pm$0.10 &	0.91$\pm$0.07&4.36$\pm$0.04&-3.33$\pm$0.65&3.46$\pm$0.25&	1.17$\pm$0.16\\
CL059\,Obj1		&	3.72$\pm$0.13 & 1.08$\pm$0.05&3.66$\pm$0.02&-0.26$\pm$0.08&3.64$\pm$0.25&	1.51$\pm$0.47\\
DBS75\,Obj1		&	5.50$\pm$0.24 &	1.35$\pm$0.04&4.57$\pm$0.01&-2.04$\pm$0.54&4.79$\pm$0.05&	1.79$\pm$0.19\\
DBS93\,Obj2		&	4.81$\pm$0.28 & 0.94$\pm$0.04&3.92$\pm$0.12&-1.40$\pm$0.37&3.35$\pm$0.16& 2.33$\pm$0.69\\
DBS100\,Obj1	&	5.10$\pm$0.18 &	0.40$\pm$0.21&4.16$\pm$0.02&-2.18$\pm$0.78&1.60$\pm$0.20&	0.02$\pm$0.71\\
\enddata
\label{sed_table}
\end{deluxetable}

\section{Summary}

In this paper we are reporting some follow-up spectroscopic observations and photometric analysis of eight young stellar clusters projected in the VVV disk area. Using the combination of spectroscopic parallax-based reddening and distance determinations with main sequence and pre-main sequence ishochrone fitting we determine the basic parameters (reddening, age, distance) of the sample clusters. The lower mass limit estimations show that all clusters are low or intermediate mass (between 110 and 1800 ${M}_{\odot}$), the slope $\Gamma$ of the obtained present-day mass functions of the clusters is close to the Kroupa IMF. Using VVV and GLIMPSE color-color cuts we have selected  a large number of YSO candidates, which are checked for variability, taking advantage of multi-epoch VVV observations. 
Fifty seven \% of the YSO candidates are found to show at least low-amplitude variability. In few cases it was possible to distinguish between YSO and AGB classification on the basis of the light curves. The SEDs of the spectroscopically confirmed YSO are determined, showing that in general these objects are massive.

\acknowledgments We gratefully acknowledge use of data from the ESO
Public Survey programme ID 179.B-2002 taken with the VISTA telescope,
and data products from the Cambridge Astronomical Survey Unit. 
Support for JB, SRA, RK, MK, MG, GR, MAF, CA-G, DM, CN, NM, PA, JA and MC is provided by the Ministry of Economy, 
Development, and Tourism's Millennium Science Initiative through grant IC120009,  
awarded to The Millennium Institute of Astrophysics, MAS. RK is supported by Fondecyt Reg. No. 1130140, SRA by Fondecyt No. 3140605. MK acknowledges the support by GEMINI-CONICYT project number No. 32130012.
MG acknowledges support from Joined Committee ESO and Government of Chile 2014.
This publication makes use of data products from the Two Micron All Sky Survey, 
which is a joint project of the University of Massachusetts and the Infrared
Processing and Analysis Center/California Institute of Technology,
funded by the National Aeronautics and Space Administration and
the National Science Foundation. This publication makes use of
data products from the Wide-field Infrared Survey Explorer, which
is a joint project of the University of California, Los Angeles, and
the Jet Propulsion Laboratory/California Institute of Technology,
funded by the National Aeronautics and Space Administration. This
work is based in part on observations made with the Spitzer Space
Telescope, which is operated by the Jet Propulsion Laboratory, 
California Institute of Technology under a contract with NASA.

 \end{document}